\documentclass[final,3p,times,onecolumn,hidelinks]{elsarticle}

\usepackage[figuresright]{rotating} 
\usepackage{setspace}
\usepackage{natbib}
 \bibpunct[, ]{(}{)}{,}{a}{}{,}%
\setcitestyle{round, comma, numbers,sort&compress, super}
\usepackage[hyphens]{url}
\usepackage{hyperref}
\usepackage{amssymb,bm,amsmath}
\usepackage[para,online,flushleft]{threeparttable}
\usepackage{caption, booktabs}
\usepackage[svgnames,table,xcdraw]{xcolor}
\usepackage[ruled,vlined]{algorithm2e}

\SetCommentSty{mycommfont}

\SetKwInput{KwInput}{Input}                
\SetKwInput{KwOutput}{Output}              
\usepackage{enumitem}
\newlist{stepss}{enumerate}{1}
\setlist[stepss, 1]{resume,leftmargin=*,label = \textbf{Step \arabic*}:}

\usepackage{multirow}
\usepackage{array}
\usepackage{diagbox}
\usepackage{longtable}
\newcolumntype{P}[1]{>{\centering\arraybackslash}p{#1}}
\usepackage[position=below]{subfig}
\usepackage[font=normalsize, labelsep=space]{caption}  
\newcolumntype{R}[1]{>{\raggedleft\arraybackslash}p{#1}}
\newcolumntype{L}[1]{>{\raggedright\arraybackslash}p{#1}}
\usepackage{siunitx}

\date{}

\begin{document}
\onehalfspacing
\begin{frontmatter}



\title{A data-driven case-based reasoning in bankruptcy prediction}

\author[mymainaddress1,mymainaddress2]{Wei Li\corref{mycorrespondingauthor}}
 \cortext[mycorrespondingauthor]{Corresponding author}
\ead{li.wei@nus.edu.sg}
\author[mymainaddress2,mymainaddress3,mymainaddress4,mymainaddress5,mymainaddress6]{Wolfgang Karl Härdle}\ead{haerdle@hu-berlin.de}

\author[mymainaddress7]{Stefan Lessmann}\ead{stefan.lessmann@hu-berlin.de}

\address[mymainaddress1]{Institute of Operations Research and Analytics, National University of Singapore, 117576 Singapore}

\address[mymainaddress2]{Humboldt-Universität zu Berlin, Blockchain Research Center, Unter den Linden 6, 10099 Berlin, Germany}

\address[mymainaddress3]{Wang Yanan Institute for Studies in Economics, Xiamen University, 422 Siming S Rd, 361005 Fujian, China}

\address[mymainaddress4]{Sim Kee Boon Institute for Financial Economics, Singapore Management University, 50 Stamford
Road, Singapore 178899}
\address[mymainaddress5]{Faculty of Mathematics and Physics, Charles University, Ke Karlovu 3, 121 16 Prague, Czech Republic}
\address[mymainaddress6]{Department of Information Management and Finance, National Yang Ming Chiao Tung University, No. 1001, Daxue Road, East Dist., Hsinchu City, Taiwan 300093, ROC.}

\address[mymainaddress7]{School of Business and Economics, Humboldt University of Berlin, Unter den Linden 6, Berlin 10099, Germany}

\begin{abstract}
There has been intensive research regarding machine learning models for predicting bankruptcy in recent years. However, the lack of interpretability limits their growth and practical implementation. This study proposes a data-driven explainable case-based reasoning (CBR) system for bankruptcy prediction. Empirical results from a comparative study show that the proposed approach performs superior to existing, alternative CBR systems and is competitive with state-of-the-art machine learning models. We also demonstrate that the asymmetrical feature similarity comparison mechanism in the proposed CBR system can effectively capture the asymmetrically distributed nature of financial attributes, such as a few companies controlling more cash than the majority, hence improving both the accuracy and explainability of predictions. In addition, we delicately examine the explainability of the CBR system in the decision-making process of bankruptcy prediction. While much research suggests a trade-off between improving prediction accuracy and explainability, our findings show a prospective research avenue in which an explainable model that thoroughly incorporates data attributes by design can reconcile the dilemma.
\end{abstract}




\end{frontmatter}


\section{Introduction}\label{S:1}
Bankruptcy refers to the circumstance in which an organization fails to satisfy its obligations, pay its due liabilities, and must consequently undergo debt restructuring or liquidation of its assets. Corporate bankruptcy is one of the main drivers of credit risk, and its economic consequences could be disastrous \citep{MAI2019743}. The amount of corporate insolvency throughout the economy may impose considerable negative societal costs and has a structural role in the propagation of recessions \citep{Bernanke1981}. Given these consequences, bankruptcy prediction is crucial to inform the debt management and loan approval processes of financial institutions. Much research has examined bankruptcy prediction models. An accurate model benefits investors, creditors, auditors and government policymakers \citep{Ding2012990}. However, obtaining accurate bankruptcy predictions is challenging. Typically, bankruptcy prediction is an imbalanced, high-dimensional, nonlinear classification problem that necessitates both precision and transparency in the prediction process \citep{MAI2019743, DASTILE2020106263}. In this study, we aim to create an automatic design case-based reasoning (CBR) system with an asymmetrical structure that can accurately predict bankruptcy and demonstrate interpretability in the decision-making process. In particular, we find that applying the asymmetrical function to calculate feature similarity is an effective way to enhance prediction accuracy by disentangling the asymmetrically distributed information existing in financial data, further boosting the economic interpretability of the optimized parameters. 

Recently, classification techniques based on machine learning (ML) have gained much popularity in the field and have demonstrated impressive predictive accuracy. Contrary to traditional approaches, ML models like neural networks, random forest, or boosting machines have fewer data-related assumptions and can capture nonlinear dependencies in the feature-target relationship in a data-driven manner. These characteristics equip ML models with more flexibility to adapt to complex data. However, a major issue with the current generation of ML algorithms is that they prioritize accuracy above explainability. Thus, they have been criticized for their black-box character, meaning that a trained model does not clarify the transformation of feature values into a (bankruptcy) prediction \citep{Cui2006597}. This opaqueness limits the applicability and value of ML in sensitive fields such as banking or healthcare. In addition, it is not straightforward how ML model predictions assist pre/post-bankruptcy decisions, such as, for example, aiding decision-makers to prevent bankruptcy a priori. Moreover, current legislation, such as the general data protection regulation (GDPR) in the European Union, further limits the application of black-box ML models. In particular, decisions based solely on automated processing are forbidden, and meaningful explanations of the rationale involved must be provided \citep{Voigt2017}. The fields of interpretable ML and explainable AI aim at addressing these challenges by developing approaches that give accurate predictions and decision-making rationales, which allow humans to comprehend and trust the models and the recommendations they generate. 

CBR is a promising approach for providing transparency and explainability in the process of bankruptcy prediction. Transparency is the explanation objective that enables people to comprehend and investigate how the calibrated model generated the forecast. It is intuitive to understand the concept of finding a solution for a new problem based on similar cases for which solutions are available. This understanding underlines the decision process in CBR \citep{Sormo2005}. The natural explanation capability of the CBR system has increased its applications in a variety of domains \citep{CHI199367, Morris1994, Deng_EJOR, OROARTY1997417, Beddoe_EJOR,  Zhuang_EJOR}. It is especially valued by decision-support systems if it is desired to comprehend how the system generates a suggestion \citep{Moxey201025}, such as when a doctor's diagnosis is being supported \citep{GUESSOUM2014267, LAMY201942}. The study of \citet{Cunningham2003122} shows that the explanation of predictions is important, and case-based explanations will significantly improve user confidence in the solution compared to rule-based explanations or only displaying the problem solution. In the context of bankruptcy prediction, reasons for a company's insolvency can be analyzed by comparing its status with other similar companies, which is helpful for developing a rescue strategy. Additionally, distinct input variables provide additional information that contributes unequally to the prediction. Identification of the most relevant facts may be utilized to adjust a company's future action choices for resolving insolvency. For instance, the determination of a merger with an insolvent company requires careful consideration. Several ML algorithms, including neural networks and random forest, have disregarded this crucial need. Meanwhile, the CBR method allows users to recognize which factors are essential for the decision-making process. 

However, the current generation of CBR systems suffers from inaccuracy in bankruptcy prediction. In the CBR mechanism, it is vital for the CBR process to get the most similar instances from the available database. Thus, a successful CBR retrieval method relies heavily on domain expertise and design experience. Even for specialists, it is difficult to gain efficient domain knowledge and establish a priori the set of most effective parameters in similarity calculation functions to solve a particular issue. For instance, it is not straightforward to design a function and assign its parameters to compare the similarity between two companies. In this research, we consider a data-driven CBR system with an asymmetrical automated design to improve the accuracy of bankruptcy predictions. Cornerstones of the underlying methodology have originally been proposed by \citet{Li20221}. The results of the extensive experimental studies show the effectiveness of the asymmetrical local similarity function in escalating the prediction performance among the best ML models, such as XGBoost, random forest and neural network models. To further enhance the explainability of model predictions and clarify their use cases, we propose Shapley-CBR, an approach to evaluate the contributions of feature values to the model-predicted bankruptcy probability. Lastly, prior work on CBR systems does not study the economic meanings behind local parameters. We comprehensively show the explainability of the parameters and enhance the transparency in the asymmetrical CBR system.

\subsection{Literature review}
The prediction of bankruptcy for companies has been an extensively researched area since the late 1960s \citep{Edward1968589}. Much of the current studies aim to develop more accurate models, which has contributed to the introduction of numerous algorithms for bankruptcy prediction \citep{MARTIN1977249, SHIN2002321, Hardle2005,  LENSBERG2006677, SUN2007738, Hardle2011}. \citet{MARTIN1977249} presented a logistic regression to predict bankruptcy based on the data obtained from the Federal Reserve System. In the study of \citet{Hardle2011}, a support vector machine was used to predict the solvency or insolvency of companies. The results showed that SVMs are capable of extracting useful information from financial data and then labeling companies by giving them score values. The study from \citet{SUN2007738} provided operational guidance for building Naive Bayes Bayesian network models for bankruptcy prediction using financial and accounting factors. In the last decade, neural network models experienced rapid development and increased their contributions to bankruptcy prediction research. \citet{HOSAKA2019287} applied convolutional neural networks to predict bankruptcy by presenting correspondence between financial ratios as grayscale images. In the study of \citet{DUJARDIN2021869}, an ensemble of self-organizing neural networks was used in bankruptcy prediction. The results from the research showed that deep neural networks significantly increase the prediction accuracy that can be obtained with conventional methods. However, much of the studies in this stream of literature work on the improvement of the accuracy and ignore the investigation of its explainability in bankruptcy-based decision makings. Distinguishing from the literature, our study primarily endeavours to develop an explainable model with high-level predictability for bankruptcy prediction. 

Especially, transparency and economic meanings are pivots in financial sectors. Several prior studies have applied the CBR system in bankruptcy prediction and business decision making \citep{JO199797, AHN2009599, VUKOVIC20128389, CHUANG2013174, INCE2014205}. \citet{JO199797} compared the bankruptcy prediction performance of multivariate discriminant analysis, CBR, and neural networks. The results showed neural networks performed best. For the CBR, a simple weighted Euclidean metric model was used to measure similarities. In the study of \citet{AHN2009599}, a hybrid CBR system using a genetic algorithm was proposed to optimize feature weights to increase bankruptcy prediction accuracy. The authors stated that the CBR system has a good explanation ability and high prediction performance over the other ML methods. However, the conclusion is ambiguous, and no experiments or case studies were conducted to substantiate the assertions. \citet{CHUANG2013174} developed three hybrid CBR methods for enhancing CBR performance in bankruptcy prediction. Although multiple experiments to compare their performance had been conducted, the study lacked comparison with other widely-used models and did not examine the explainability of CBR in bankruptcy prediction. In the study of \citet{VUKOVIC20128389}, a CBR method combined with an optimization algorithm was used for credit scoring. The results showed that the proposed CBR method improved the performance of the traditional CBR system and outperformed the $k$-nearest neighbour classifier. However, the study did not mention the explainability of the CBR method. The recent study of \citet{Li20221} provided abundant empirical evidence that the asymmetrical CBR system can predict as accurately as the majority of ML models. However, the research did not examine its efficacy in predicting insolvency. Moreover, the lack of explainability in parameters still needed supplementation. Compared with the previous research, our study aims to examine and explore the transparency and explainability of CBR comprehensively. 

Besides, in light of the significant drawback of black-box ML model implementations in the financial sector, a number of recent studies have focused on post-hoc approaches for explicating ML models. The most prevailing one is Shapley value, which is widely used in cooperative game theory, allocating the entire rewards to the players on the assumption that they are all cooperating \citep{Hart2017}. The value is the average marginal contribution of a player across all possible coalitions. The Shapley method has been applied in numerous fields, such as minimal cost spanning trees \citep{Bergantinos2008}, risk capital allocation \citep{BALOG2017614}, closed-loop supply chain \citep{ZHENG2019227} and liability games \citep{CSOKA2022378}. In recent years, its application has been extended to other sectors, e.g., improvement in the explainability of ML models \citep{Lundberg2017}. \citet{Lundberg2020} investigated the explainability of the Shapley method in tree-based ML models, such as random forest, decision trees and gradient boosted trees. \citet{Li2021} applied the Shapley values to explain the predictive results of the support vector machine model for electricity price forecasting under consideration of market coupling. In the study of \citet{Jabeur2021}, the Shapley interaction values are used to explain the forecasting of the gold price with the XGBoost algorithm. The conservative business and banking sector are often resistive to revolutionary algorithms. It has raised the question that if the black-box model is explainable. The black-box nature of neural networks constrains their explainability in the results. According to \citet{Rudin2019}, the applications of black-box ML models are criticized for high-stakes decision-making throughout society. This study employed Shapley value to enhance the explainable CBR system by evaluating the feature importance in the prediction probability of insolvency and providing decision support for rescuing an insolvent company.

The contribution of this article is threefold. Our method is first to consider utilising the asymmetrical nature of financial features to design and improve the CBR system. In particular, we examine the efficiency of the asymmetrical CBR system in solving bankruptcy prediction problems, which performs significantly better than the conventional distance-based CBR system and can perform as well as the widely used ML models, such as XGBoost and neural networks models. The reason behind this fact is that the asymmetrical local similarity function can better capture the asymmetrical properties that exist in financial features compared with existing other similarity functions. Our study provided a new research direction that a data-driven well-fitted model can not only escalate the prediction accuracy but enhance explainability. This can overcome the conflict between model accuracy and model intractability inaugurated by \citet{BARREDOARRIETA202082}. Second, according to the study of \citet{BARREDOARRIETA202082}, transparent ML models convey some degree of explainability by themselves. The levels of transparency can be evaluated based on three aspects, i.e., algorithmic transparency, decomposability and simulatability. Our study aims to examine and explore the transparency and explainability of CBR comprehensively in terms of the three aspects. In particular, we first illustrate that the fitted polynomial parameters in the proposed function can reveal the economic meanings for the similarity comparison, which enhance the algorithmic transparency and decomposability of the CBR system. It is useful for users to investigate the reasons behind bankruptcy. Third, the posterior probability estimator of CBR combined with a post-hoc explainability technique, Shapley value, is first developed for investigating the influence of features on the predicted insolvent probability. The applications of the Shapley-CBR concerned with bankruptcy-related decisions have been extensively investigated in empirical studies. 

The rest of this paper is organized as follows. Section 2 is the data description. Section 3 introduces the asymmetrical CBR system. It is followed by the experiment design, introduced in Section 4. The experiment results and explanation instances are shown in Section 5 and Section 6. Section 7 concludes the paper. 

\section{Data description}
The dataset applied in this study comes from the credit reform database provided by the Blockchain Research Center (BRC, \url{https://blockchain-research-center.de/}). It contains financial information from 20,000 solvent and 1,000 insolvent German companies. The last annual financial report of a company is used to predict its solvency in the next year. There are 28 variables, i.e., cash, inventories, equity, and current assets. The list of features is shown in Table \ref{t:features}.
\begin{table}[ht!]
\caption{Features table.}\label{t:features}
 \begin{threeparttable}
 \begin{tabular*}{\textwidth}{p{2cm}p{6cm} p{2cm} p{6cm}  } 
\toprule
\toprule
No.&Feature&No.& Feature\\
\midrule
VAR1&Cash &VAR15&Accounts payable (A.P.)  \\
VAR2&Inventories& VAR16&Sales\\
VAR3&Current assets&VAR17&Administrative expenses\\
VAR4&Tangible assets&VAR18 &Amortization depreciation \\
VAR5&Intangible assets& VAR19&Interest expenses\\
VAR6&Total assets&VAR20&EBIT  \\
VAR7&Accounts receivable (A.R.)&VAR21&Operating income \\
VAR8&Lands and buildings&VAR22&Net income \\
VAR9&Equity&VAR23&Increase inventories\\
VAR10&Shareholder loan&VAR24&Increase liabilities \\
VAR11&Accrual for pension liabilities&VAR25&Increase cash \\
VAR12&Total current liabilities&VAR26&A.R. against affiliated companies\\
VAR13& Total long-term liabilities &VAR27&A.P. against affiliated companies\\
VAR14&Bank debt&VAR28& Number of employees\\
\bottomrule
\bottomrule
\end{tabular*}
 \begin{tablenotes}
    \end{tablenotes}
  \end{threeparttable}
\end{table}

\begin{centering}
\begin{figure}[ht!]
\caption{Two dimensional PCA and UMAP projection of the dataset.}
\includegraphics[width=\textwidth]{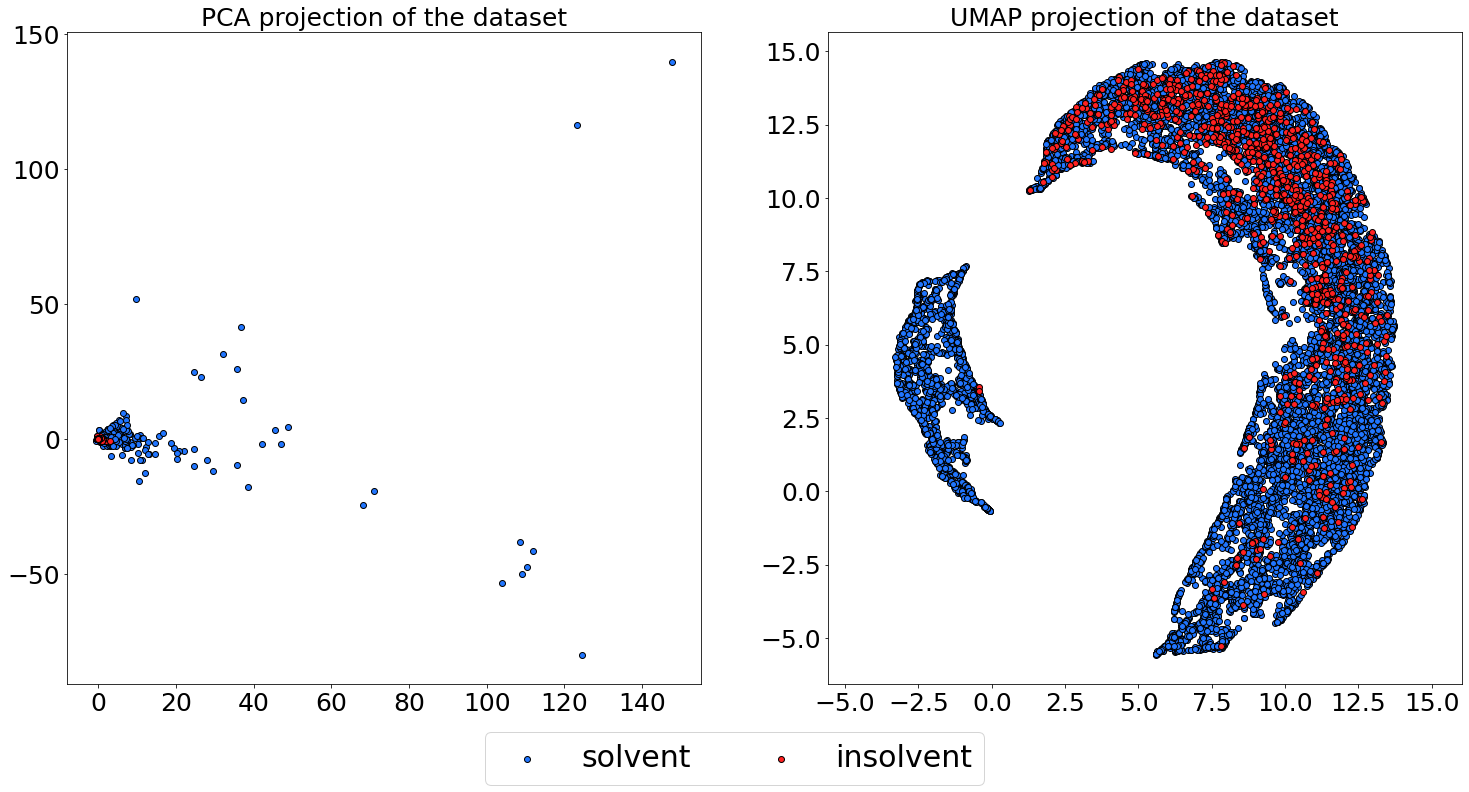}
\label{f:data}
\end{figure} 
\end{centering}

The dataset has been visualized in Figure \ref{f:data}, based on two-dimensionality reduction techniques, Principal Component Analysis (PCA), and Uniform Manifold Approximation and Projection (UMAP). From the PCA plot in Figure \ref{f:data}, we can see that the insolvent data points are agglomerated together, but clearly not enough to set them apart from the solvent ones. From the UMAP plot in Figure \ref{f:data}, the solvent data points are clearly separated into two groups. The fiscal year of companies in the small group was 2003 when Germany was experiencing a temporal recession, and the German government announced a set of policies to spur the economy \citep{Dustmann2014167}. This is one reason that the small group is isolated from the main group. However, the insolvent data points are scattered, and there are no big clusters separating the sign sufficiently. 

\begin{centering}
\begin{figure}[h!]
\centering
\caption{The box plot of feature variables.}
\includegraphics[width=\textwidth]{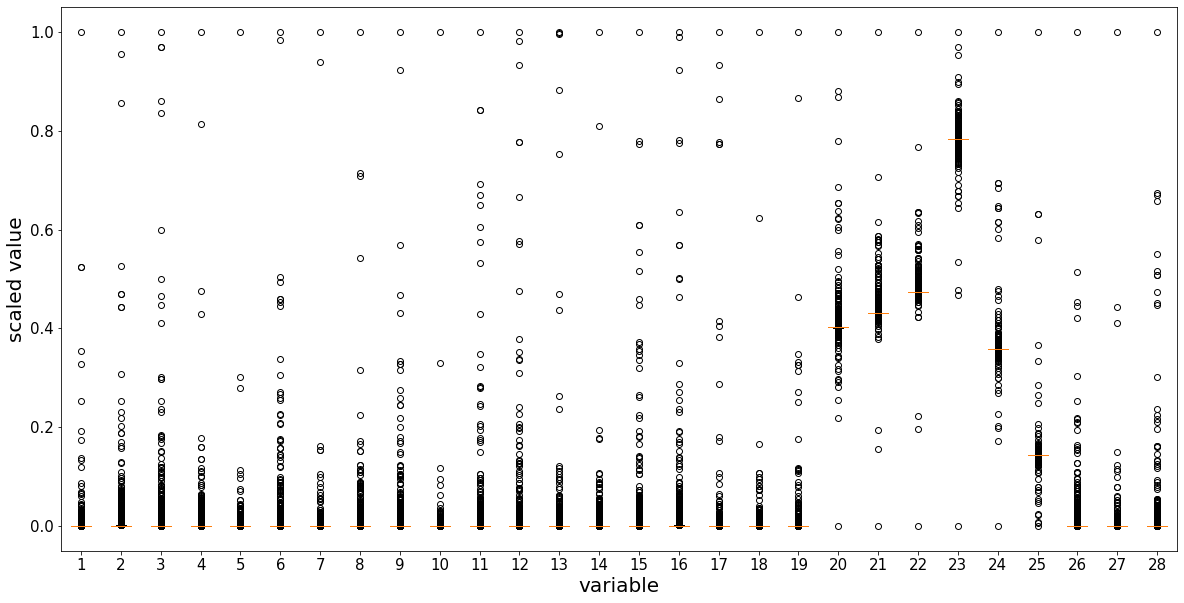}
\caption*{\small{\textit{Notes.} The low, median and high percentiles are 25\%, 50\% and 75\%, respectively. The whiskers of the boxplot are a representation of the multiple 1.5 of the innerquartile range. The value of each variable has been scaled between 0 and 1. Those variables are asymmetrically distributed, and the majority of them are centralized.}}
\label{f:var_plot}
\end{figure} 
\end{centering}

The scaled value (between 0 and 1) of each feature variable has been shown in Figure \ref{f:var_plot}. From the figure, we can observe that those features are asymmetrically distributed. Such distributional property is common for financial data features. For instance, the majority of companies are small and medium-size, which often suffer from limited cash flow. By contrast, few are large companies that can dispose of large amounts of cash. Thus, it is reasonable to investigate the effectiveness of the asymmetrical similarity function when calculating similarities of cash or other features between a query company and its relatively smaller and larger companies, respectively.

\section{Methodology}\label{sec:method}
CBR is a typical explainable ML approach, the decision-making process of which is analogous to that of humans, i.e., a solution to unravel new problems can be considered based on past experiences. The logic framework of CBR typically includes four \textit{R} steps \citep{Aamodt1994}: 1) To solve a new issue without a solution, previous cases with solutions are \textit{retrieved (R)} to determine which are most similar to the current problem; 2) The solutions of the most similar problems are \textit{reused (R)} to offer a solution for the new case. 3) Whether the new issue has been resolved can be \textit{revised (R)}; 4) The resolved new problem will be \textit{retained (R)} in the case database for future use. Clearly, the most crucial of the four \textit{R} steps is the first \textit{R}, which entails retrieving the case database and identifying the most similar cases, as this influences the success of the subsequent decision-making process. By calculating the similarity of cases, it can be mathematically converted into a prediction problem.

\subsection{Similarity calculation in CBR}
CBR systems aim to find the most similar cases of a query case in the case base (dataset) and detect its solutions based on similar cases. In most cases, the similarity computation adheres to the local-global principle for customizing the similarity measure for each attribute and constructing a knowledge model, which is prevalent in attribute-based CBR systems \citep{Richter2013}. Given a query  case $Q$ and a case $C$ from a $L$-dimensional database ($L$ features), the distance between two cases on the $j$th feature can be employed to measure the local similarity, which can be expressed as $\mathrm{dist}_j(q_j,c_j) = | q_j - c_j |$. The Euclidean distance metric has been commonly applied for distance-based CBR similarity measure in the area of financial distress prediction \citep{LI200910085}. Let $\mathrm{Sim}(Q,C)$ express the global similarity between case $Q$ and case $C$, and $\bm{w_j}$ be feature weights ($\sum \bm{w_j} =1$). The mechanism for measuring similarity in Euclidean space could be derived from the distances between two cases on each feature, as denoted by ECBR and shown below:

\begin{equation}
   \mathrm{Sim}(Q,C) =\frac{1}{1+\sqrt{\sum_{j=1}^L {\left[\bm{w_j}\times \mathrm{dist}_j(q_j,c_j)\right]}^2}}
\end{equation}

The Manhattan distance metric is another extensively used distance-based measure of similarity in the field of CBR (MCBR). It could be demonstrated as follows:

\begin{equation}
   \mathrm{Sim}(Q,C) =\frac{1}{1+\sum_{j=1}^L \left[\bm{w_j}\times \mathrm{dist}_j(q_j,c_j)\right]}
\end{equation}

In addition, the grey coefficient degrees have been developed for local similarity calculation (GCBR). Let $\hat{C}$ expresses the case base (dataset) without case $C$, and $\mathrm{sup}| q_j - \hat{c}_j|$ and $\mathrm{inf}| q_j - \hat{c}_j|$ represent the minimum and maximum distances of case $Q$ and the case base on the $j$th feature, respectively. The similarity between two cases on the foundation of weighted average of grey coefficient degrees can be calculated as follows:
\begin{equation}
   \mathrm{Sim}(Q,C) =\sum_{j=1}^L {\left[\bm{w_j} \times \mathrm{degree}_j(q_j,c_j)\right]}^2
\end{equation}
\noindent where
\begin{equation}
\mathrm{degree}_j(q_j,c_j) = \frac{2\times \mathrm{inf}| q_j - \hat{c}_j|+\mathrm{sup}| q_j - \hat{c}_j|}{2\times \mathrm{dist}_j(q_j,c_j) + \mathrm{sup}| q_j - \hat{c}_j|}
\end{equation}
\subsubsection{Asymmetrical CBR}

The design of the asymmetrical CBR (ACBR) also follows the local-global principle. In particular, the global similarity is measured by the square root of the weighted sum of the asymmetrical local similarities. The similarity between $Q$ and $C$ can be described as follows:
\begin{equation}\label{eq1}
   \mathrm{Sim}(Q,C) = \sqrt{\sum_{j=1}^L \bm{w_j}\times \left[\mathrm{sim}_j(q_j,c_j)\right]^2}
\end{equation}
\noindent where, for the attribute $j$, $\mathrm{sim}_j$ is the local similarity function, $q_j$ and $c_j$ are attribute values from the case $Q$ and $C$, respectively. $\bm{w_j}$ stands for the weight (global parameters) of the attribute $j$. For the local (feature) similarity, asymmetrical polynomial functions are commonly used to measure the similarity of attribute-value \citep{Bach201217}. It can be represented as:
  \begin{equation}\label{eq2}
   \mathrm{sim}_j(q_j,c_j)=
    \begin{cases}
      \left[\frac{D_{j}-(c_j-q_j)}{D_{j}}\right]^{\bm{a_j}}, & \text{if}\ q_j \leq c_j \\
      \left[\frac{D_{j}-(q_j-c_j)}{D_{j}}\right]^{\bm{b_j}}, & \text{if}\ q_j > c_j
    \end{cases}
  \end{equation}
  
\noindent where $D_{j}$ stands for the difference between the maximum and minimum value of attribute $j$ in the dataset. $\bm{a_j}$ and $\bm{b_j}$ are the degree (local parameters) of polynomial functions. The similarities between a query case and all the cases in the dataset are calculated based on Equations (\ref{eq1}) and (\ref{eq2}). Thus, the global parameter $\bm{w_j}$ and local parameters $\bm{a_j}$ and $\bm{b_j}$ are required to automatically design a data-driven CBR system without human involvement. The following subsection is to introduce how to obtain those parameters. A simple instance of the similarity calculation can be found in Appendix A.

\subsection{Data-driven ACBR system design}
In addition to the parameters $\bm{w_j}$, $\bm{a_j}$ and $\bm{b_j}$, the ACBR system classifies a given query case by picking the $K$ most similar examples from the matching dataset. Then, majority voting is used to give the query case the class with the highest frequency among similar cases. Thus, the $K$ is the other parameter to be determined. The well-known $k$-nearest neighbors method ($K$-NN), which may be termed a non-parametric CBR, is used to find the parameter $K$. In the $K$-NN paradigm, a case is typically categorized by the majority vote of its $K$ distance-based neighbors. Indeed, the only parameter affecting the classification accuracy of the $K$-NN model that must be calculated is $K$. Cross-validated grid search across a parameter grid may be used to determine the best $K$ for a given dataset.

\begin{centering}
\begin{figure}[!ht]
\caption{Illustration of the cost function.}
\centering
\includegraphics[width=\textwidth]{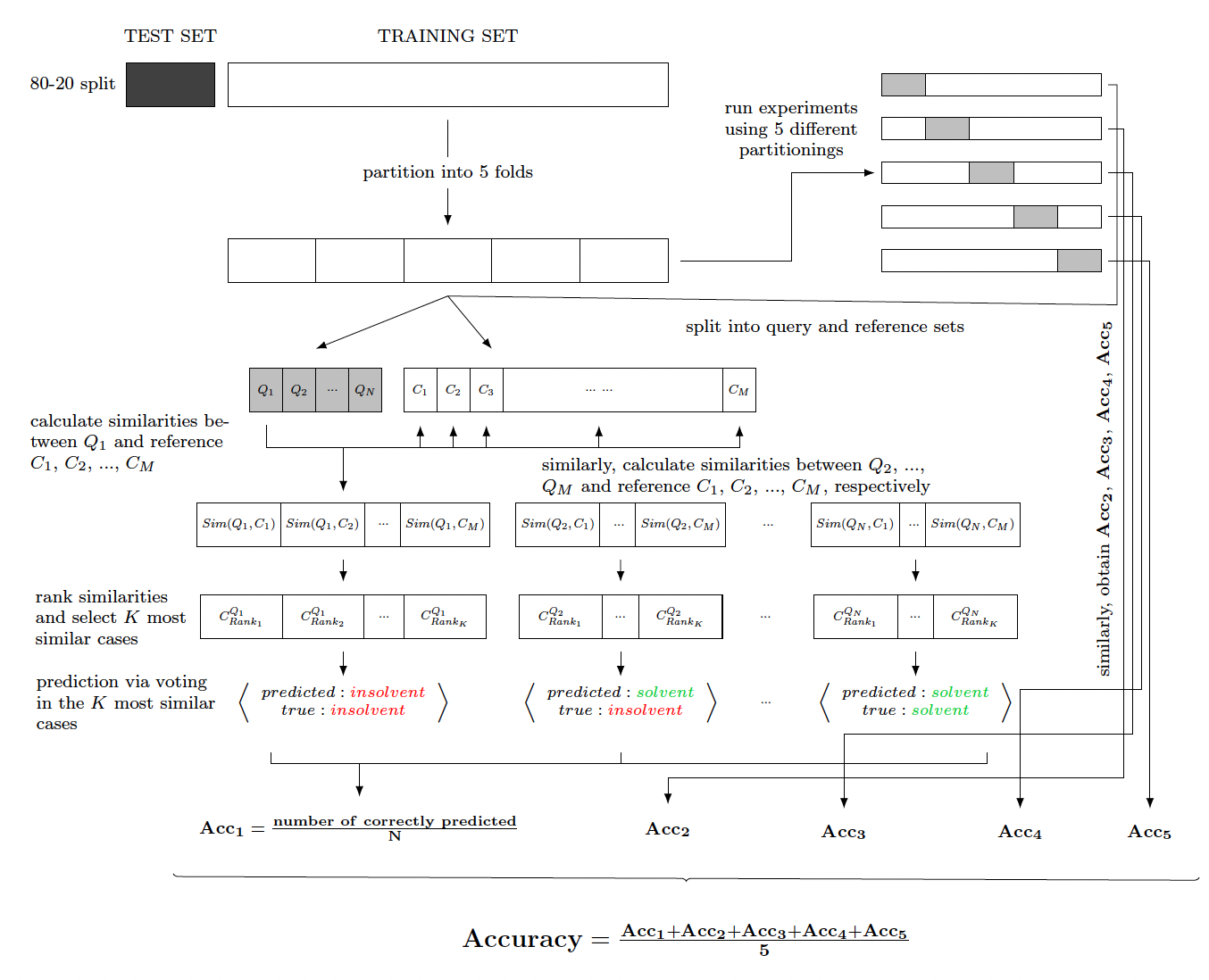}
\caption*{\small{\textit{Notes.} The cost function is the average prediction accuracy calculated based a 5-fold cross-validation, used for the prediction performance evaluation of a CBR system $\mathrm{Sim}(Q,C)$ with specific parameters ($K$, $w_j$, $a_j$, $b_j$). The training set is partitioned into 5 folds (one for query and four for reference sets) to run 5 experiments. For each experiment, each $Q_n$ ($n$ = 1, ..., $N$) from the query set will be used to calculate its similarities with all $C_m$ ($m$ = 1, ..., $M$) from reference set. For instance, for $Q_1$, the similarities $\mathrm{Sim}(Q_1,C_1)$, $\mathrm{Sim}(Q_1,C_2)$, ..., $\mathrm{Sim}(Q_1,C_M)$ will be calculated. Then, those similarities will be ranked, and the $K$ cases $C_{Rank_1}^{Q_1}$, $C_{Rank_2}^{Q_1}$, ..., $C_{Rank_K}^{Q_1}$ with highest similarities values will be selected. The prediction for $Q_1$ will be made based on the voting in the labels of those $K$ cases. Similarly, the predictions for $Q_2$, $Q_3$, ... and $Q_N$ can be obtained. Afterwards, the accuracy of the prediction in this experiment $\bf{Acc_1}$ can be calculated, which is equal to ${\text{\bf{number of correctly predicted}}/\bf{N}}$. Similarly, $\bf{Acc_2}$, $\bf{Acc_3}$, $\bf{Acc_4}$ and $\bf{Acc_5}$ for the other experiments can be computed. Last, the prediction performance for the CBR with the specific parameters can be evaluated by $\text{\bf{Accuracy}}$, which is equal to $\bf{(Acc_1 + Acc_2 + Acc_3 + Acc_4 + Acc_5)/5}$. }}
\label{f:5-fold}
\end{figure} 
\end{centering}

To obtain the $\bm{w_j}$, feature importance scoring methods are employed. Feature importance scoring is a task that calculates the relevance between the descriptive features and the target feature, which is commonly used in ML model design \citep{Zien2009}. In this study, six scoring methods are applied to generate six sets of global weights $\bm{w_j}$, which are Gini \citep{Ceriani2012}, Information entropy \citep{Kullback59}, Mutual information \citep{Kraskov2004}, Chi2 \citep{Cost1993}, ANOVA \citep{LIN201164} and ReliefF \citep{Kononenko1997}. The description of the methods can be found in Appendix B. Consequently, we create six CBR systems based on the generated weights. For each created CBR, the optimal local parameters, $\bm{a_j}$ and $\bm{b_j}$ of polynomial functions, will be searched and evaluated to find the optimal CBR system structure. However, the loss function is non-differentiable, which is an average accuracy calculated via 5-fold cross-validation. The illustration of the cost function can be seen in Figure \ref{f:5-fold}.

Thus, particle swarm optimization (PSO) is applied for searching the optimal local parameters since the algorithm does not require that the optimization problem be differentiable as is required by classic optimization methods such as gradient descent used in training neural networks. PSO is a commonly used evolutionary algorithm. PSO is superior to other popular algorithms, such as genetic algorithms, in terms of efficiency and iteration while determining the optimal solution \citep{Wihartiko2018}. Its popularity has increased because to its several advantages, which include resilience, efficiency, and simplicity. The other challenge is that the training process requires enormous calculations. Thus, parallel computing with GPU has been used to speed up the computation since the similarity calculation of cases is independent of each other and can be proceeded simultaneously. As we can see in Figure \ref{f:5-fold}, the calculation of [$\mathrm{Sim}(Q_1,C_1)$, $\mathrm{Sim}(Q_1,C_2)$, ..., $\mathrm{Sim}(Q_N,C_M)$] can be conducted parallelly. The parallel computing for the similarity calculation is explained in Appendix C. The explanation of PSO can be found in Appendix D.

Finally, after evaluating the performance of the six designed ACBR systems through cross-validation, the best-validated one will be selected and used for bankruptcy prediction. The designing process of the proposed ACBR can be described as follows:

\begin{algorithm}[!h]
  
\KwInput{Financial data input}
\KwOutput{Designed ACBR system}
Data processing. \\
Determine the number of the most similar cases $K$ for retrieval with $K$-NN algorithm.\\
\While{There are more feature scoring methods}
  {
    Score the features and assign the weights $\bm{w_j}$ using the feature importance scoring method. \\
    Optimize the parameters, $\bm{a_j}$ and $\bm{b_j}$, of the local similarity functions using PSO algorithm. \\
    Evaluate the CBR system via cross-validation.\\ 
    }
Compared the performance of all the trained CBR systems, select the best-validated one. \\
\caption{Data-driven ACBR system design}
\end{algorithm}
\vspace{2mm}


\subsection{Probability of forecasting} \label{Probability}
A model is necessary to provide the likelihood of an event's occurrence, not merely its anticipated categorization. It is crucial to convey the posterior probability in order to boost user trust in the CBR system's answer \citep{Li20221}. In particular, bankruptcy prediction is a binary classification problem with classes $Y$ ($Y=0$ (solvent) or $Y=1$ (insolvent)). Assume a dataset $X$ includes $N$ labeled cases $x(n)$, $n = 1, ..., N$, and for a query case $x$, $K'$ is the number in the $K$ most similar cases belong to the class insolvent ($Y = 1$), the estimate of the insolvent probability $\hat{P}(Y=1|X=x)$ is given by: 
\begin{equation}
\hat{P}(Y = 1|X = x) = \frac{K'}{K}
\end{equation}
However, it defies common sense to assume that each instance in the $K$ most comparable situations is given the same weight. The probability computation should favour the more similar scenario over the less similar case. Therefore, it is preferable to generalize this estimator by giving various probabilities to various related cases. Let the probabilities assigned to the $K$ most similar cases be $p_1$, ..., $p_K$ and the label $B_i=1$ if the $i${th} case belongs to the class $Y=1$ and $B_i=0$ otherwise. These probabilities are greater than or equal zero, monotonically decreasing, and sum to 1: $\sum_{i=1}^Kp_i=1$ (constraints). Then the probability estimate of the insolvent is given: 
\begin{equation}
\hat{P}(Y = 1|X = x) = \sum_{i=1}^{K+1}  B_i \times p_i
\end{equation}
The probabilities $p_1$, ..., $p_{K+1}$ are determined by maximizing the likelihood of the dataset $X$. It is worth to note that the $K+1$ probabilities rather than $K$ probabilities are used. The $B_{K+1} \times p_{K+1}$ is a regularization term to prevent obtaining $-\infty$ log likelihood by assigning $B_{K+1}=1/2$. Further, to reduce the constraints when optimizing log likelihood function to obtain the probabilities, a softmax representation is used:  
\begin{equation}
p_i = \frac{e^{\omega_i}}{\sum_{j=1}^{K+1}e^{\omega_j}},\;\; \text{ for } i = 1 ,..., K+1
\end{equation}
where the parameters $\omega_i$ can be any value and constrained to be monotonically decreasing. Then, the estimate function of the insolvent probability becomes:  
\begin{equation}
\hat{P}(Y=1|X = x) = \frac{\sum_{i=1}^{K+1}B_{i}e^{\omega_i}}{\sum_{j=1}^{K+1}e^{\omega_j}}
\end{equation}
Let $B(n)$ denote the class membership of $x(n)$. The likelihood $\mathcal{L}$ of the $N$ cases dataset $X$ is: 
\begin{equation}
\mathcal{L}= \prod_{n=1}^N \hat{P}(Y=1|X=x(n)) =  \prod_{n=1}^N \left[\frac{\sum_{i=1}^{K+1}B_i(n)e^{\omega_i}}{\sum_{j=1}^{K+1}e^{\omega_j}}\right]
\end{equation}
where the different probability estimates are assumed to be independent as the dependent case is complicated to analyze. Finally, the log likelihood is given:
\begin{equation}
\log(\mathcal{L}) = \sum_{n=1}^N\log\left[\frac{\sum_{i=1}^{K+1}B_i(n)e^{\omega_i}}{\sum_{j=1}^{K+1}e^{\omega_j}}\right]
\end{equation}
subject to the constraint:
\begin{equation*}
  \omega_1 \geq \omega_2 \geq \omega_3\geq ... \geq \omega_K
\end{equation*}
The weighting parameters $\omega_j$ are determined by maximizing the log likelihood function.

\subsubsection{Learning from prediction probability}
The prediction probability provided by the CBR system, a number between zero and one, can be considered as an ex-ante bankruptcy risk signal, for instance, the creditworthiness of a company. In particular, the annual financial statement of companies can be used to calculate the scores of solvency in subsequent fiscal years, which can support some decision-making processes. 

In addition, some insolvent companies are successfully merged by other companies as a form of an exit strategy, and some others invite new investors or sell liquid assets to get through financial distress, but others failed \citep{NISHIHARA20211017}. It is important to consider the reorganization strategy before adopting a decision on a distressed firm's options \citep{Cappelen20192832}. To assist the investment and financing decisions, the prediction confidence in CBR can be used as an ex-post metric to examine a firm's choice. In particular, the determination can be evaluated in terms of increasing the score of solvency. For instance, the efficiency of a capital injection can be measured based on the improvement of solvency confidence. Further, Shapley values are employed to show the contributions of financial variables to the improvement. 

\subsection{Shapley-CBR}
Shapley-CBR is used to evaluate the feature importance for influencing the prediction probability of the CBR system. Shapley values are originally introduced as a pay-off concept from cooperative game theory \citep{Lundberg2017}. When referring to ML models, the notion of pay-off corresponds to the feature importance for predictive models in the presence of multicollinearity. This technique is necessary to retrieve the model using all feature subsets $S\subseteq F$, where $F$ is the set of all features. Each feature is given a significance value, which indicates how that feature's inclusion will affect the model's prediction. A model $f_{S\cap\{j\}}$ and a model $f_{S}$ trained with that feature present and withheld, respectively, are used to calculate the effect. Then, given the current input $x$, the pay-offs are computed as 
\begin{equation}\label{eq:payoff}
   \text{payoff}(x_S^j) = f_{S\cap\{j\}}(x_{S\cap\{j\}})-f_{S}(x_{S})
\end{equation}

The effects of excluding a feature depend on other features in the model, therefore the aforementioned differences are calculated for all potential subsets $S\subseteq F\setminus \{j\}$. After that, the Shapley values are calculated and applied as feature attributions. For a feature $j$, the Shapley value can be presented as a weighted average of all possible differences: 
\begin{equation}\label{eq:shapley}
    \phi_j=\sum_{S \subseteq F \setminus
\{j\}} \frac{|S|!\; (|F|-|S|-1)!}{|F|!}\{f_{S\cap\{j\}}(x_{S\cap\{j\}})-f_{S}(x_{S})\}
\end{equation}
As we are dealing with a binary response variable representing the active and default state of the companies, the term $\hat{Y}_{S\cap\{j\}}$ and $\hat{Y}_{S}$ can be rewritten as the predicted probabilities of the default $\hat{P}_{S\cap\{j\}}$ and $\hat{P}_{S}$, when using the CBR model that includes the explanatory variable $x_S^j$ versus when using the CBR model that does not include the explanatory variable $x_S^j$. Thus, Equation (\ref{eq:payoff}) becomes
\begin{equation}
   \text{payoff}(x_S^j) = \hat{P}_{S\cap\{j\}}(x_{S\cap\{j\}})-\hat{P}_{S}(x_{S})
\end{equation}
It is worth noting that Shapley-CBR presents as an agnostic explainable AI concept that can be applied to the prediction probability irrespective of the model and data that generated it.

\section{Experiment}
In this section, we describe the experimental study's principles and procedures. This section comprises five subsections. First, it is commonplace to observe missing values in financial data. To address the issue, we explain the basic concept of the CBR system regarding missing value. Next, data processing for model training is presented. It is then followed by a description of the performance measures used. Then, a list of benchmark models follows. Finally, we demonstrate the experimental design procedure.

\subsection{Prediction with missing value}
Missing values are always a pain in the neck. To control this obstacle in practice, some classification methods choose to remove all variables and observations which contain missings. This listwise deletion may lead to information loss, and new observations with missings are not predictable. Other methods solve the problem by using imputation. However, the process of replacing missing data with substituted values may introduce bias or affect the representativeness of the results. In our empirical study of evaluating models' prediction performance, listwise deletion for missing data is implemented. 

CBR is a robust and comparatively easy to handle technique, providing good performance with missing data \citep{LOW2019103127}. A value $[0, 1]$ can be assigned as a feature similarity of two cases. For instance, for any local feature similarity $\mathrm{sim}_j(q_j,c_j)$ calculation, if there are missings ($q_j$ is missing, $c_j$ is missing or both $q_j$ and $c_j$ are missing), the $\mathrm{sim}_j(q_j,c_j)$ is simply assigned to a value, such as 0.1. This characteristic enables the constructed CBR based on the existing dataset to easily adapt to any new coming cases. In our study, when there are missings, we assign the $\mathrm{sim}_j(q_j,c_j)$ to 0.

\subsection{Data processing}
The application of re-sampling strategies to obtain a more balanced data distribution is an effective solution to the imbalance problem \citep{Branco2016}. The random under-sampling method is applied when training models in this study. This method involves randomly selecting cases from the majority class and removing them from the training dataset until a balanced distribution of classes is reached. 

\subsection{Performance measure}
The evaluation of learned models is utmost important. Accuracy is the most frequently used gauge for bankruptcy model evaluation, which is the percentage of correctly classified individuals. Further, the misclassification cost aims to measure to what extent financial sectors can tolerate the misclassification results. The cost is the result of two types of errors: type I errors, in which a successful business is incorrectly labeled as failing, and type II errors, in which a failed company is misclassified as healthy. The false positive (FPR) and false negative rates (FNR) measure the probability of type I and II errors, respectively. The complementary metrics of FPR and FNR are true positive (TNR = 1 - FPR, sensitivity) and true negative rates  (TPR = 1 - FNR, specificity), respectively. Compared with accuracy, the F-measure is a more helpful indicator for determining how well a prediction model performs for its users \citep{Branco2016}. When both FPR and FNR are low, the value of the F-measure is high. G-means computes the geometric mean of the accuracy of the two classes, the value of which is maximal when the balance of accuracy of FPR and FNR is achieved. In addition, AUC is a popular alternative measure of overall discriminatory power for prediction models \citep{BRADLEY19971145}. This metric has been widely used for evaluating bankruptcy prediction models \citep{TIAN201589, TIAN2017510, Nagel20192421}. It is a more flexible performance measure because it provides a synthetic estimation of model accuracy that is totally independent of the proportion of classes and from any misclassification cost matrix \citep{DUJARDIN2021869}. Besides, considering that the solvent and insolvent companies are of very different sizes, a balanced measure, Matthews correlation coefficient (MCC), is applied to evaluate the performance of the classifiers. The description of all the listed metrics can be found in Appendix E. 

\subsection{Benchmark models}
In order to evaluate the efficiency and fidelity of the ACBR system, the experimental research employs three sets of benchmark models. First, the ECBR, MCBR, GCBR, EWCBR (Equally-weighted CBR), and EPCBR (Equally-polynomial CBR) are used as benchmarks to examine the efficiency of the asymmetrical local similarity function. In addition to comparing ACBR with the three distance-based CBR mentioned in Section \ref{sec:method}, we also involve two asymmetrical CBR models, i.e., EWCBR and EPCBR, in this experiment. In particular, EWCBR treats the features with globally equal importance (\bm{$w_j$} = 1/$K$) and locally linear related (\bm{$a_j$} = 1 and \bm{$b_j$} = 1) when constructing the CBR system. EPCBR adopts the optimally learned weights $\bm{w_j}$ using the feature importance scoring methods while taking into account the local parameters equally (\bm{$a_j$} = 1 and \bm{$b_j$} = 1). Second, eight well-known ML classifiers are used as benchmark models to assess the accuracy of ACBR system, namely logistic regression (LR), $k$-nearest neighbor ($k$-NN), decision tree (DT), support vector machine (SVM), Gaussian Naive Bayes (GNB), and multilayer perceptrons (MLP), Random Forest (RF) and XGBoost (XGB). The introduction of the benchmark models can be found in Appendix F. Third, we also investigate the efficiency of the different training metrics, i.e., Accuracy, AUC, F-measure, G-means and MCC.


\subsection{Experimental design}
As illustrated in Figure \ref{f:5-fold}, the dataset is portioned into a train and test set, with an 80-20 split. The split is made by preserving the percentage of samples for each class. Then, the 5-fold cross-validated grid-search is used to optimize the models. Last, the out-of-sample performances are evaluated in terms of multiple measure metrics with the test dataset. 

\section{Results}
This section shows the results of the out-of-sample prediction in the experiment to evaluate the efficiency and fidelity of the proposed CBR system. Specifically, we examine the effectiveness of the asymmetrical similarity function in comparison to others. Then, we compare the out-of-sample performance of the proposed CBR system to that of the most popular ML models. Finally, the accuracy of the proposed CBR system using several training metrics is investigated.

\subsection{Out-of-sample performance comparison based on different local similarity functions}
The classification results of the ACBR and the other five CBR benchmarks have been evaluated by the five measure metrics aforementioned, as shown in Table \ref{t:scores_local}. From Table \ref{t:scores_local}, we can observe that the ACBR performs overwhelmingly better than the other CBR systems. Particularly, the naïve benchmark EWCBR did not outperform the distance-based CBR as anticipated. In addition, we discover that EPCBR marginally improved its predictive performance by utilizing the learnt weights, but that the information provided by the global features is insufficient to improve the CBR system's predictability considerably. In the meanwhile, the exceptional performance of the ACBR system demonstrates the efficacy of the asymmetrical similarity function, which can capture the prevalent asymmetrical characteristics in financial data characteristics.

\begin{table}[ht!]
\caption{Out-of-sample performance calculated based on different local similarity functions.}\label{t:scores_local}
 \begin{threeparttable}

 \begin{tabular*}{0.99\textwidth}{p{2.cm} P{2.5cm}P{2.5cm}P{2.5cm}P{2.5cm}P{2.5cm}} 
\toprule
\toprule
&{Accuracy}&{AUC}& F-measure&G-means&MCC \\

\midrule
ECBR&0.6675&0.7250&0.7483&0.7219&0.2389\\
MCBR&0.6823&0.7126&0.7595& 0.7118&0.2286\\
GCBR&0.6203&0.6848&0.7115&0.6807&0.1926\\
EWCBR&0.6672&0.7101& 0.7481&0.7083&0.2236\\
EPCBR&0.6826&0.7128&0.7597&0.7119&0.2288\\
ACBR&\textbf{0.7528}&\textbf{0.7358}&\textbf{0.8106}&\textbf{0.7355}&\textbf{0.2704}\\

\bottomrule
\bottomrule
\end{tabular*}
 \begin{tablenotes}
 \normalsize
      \item  \small{\textit{Notes.} The best values of each column are depicted in bold.}
    \end{tablenotes}
  \end{threeparttable}
\end{table}
\subsection{Out-of-sample performance of different ML models}
The out-of-sample prediction results of the ACBR and eight benchmarks are shown in Table \ref{t:scores_performance}. From Table \ref{t:scores_performance}, we can observe that no classifier has the best performance across all measures. The finding is consistent with the results of research that conducted a similar experiment to evaluate ML models using a variety of metric measurements \citep{Novakovic2016}. In addition, the linear-based models exhibit impoverished prediction performance. The result is aligned with observations for insolvent/solvent classes in Figure \ref{f:data}, which are intuitively difficult to be distinguished by linear methods. The XGB performs the best among all models, followed by the  ACBR, RF, and MLP, which all perform much better than the other models based on a number of measurement measures. Nonetheless, the MLP, XGB and RF did not indiscriminate their performance from the proposed CBR, indicating that the explainable model is capable of delivering a high degree of accuracy.
\begin{table}[ht!]
    \caption{Out-of-sample prediction results.}\label{t:scores_performance}
 \begin{threeparttable}

 \begin{tabular*}{0.99\textwidth}{p{2cm} P{2.5cm}P{2.5cm}P{2.5cm}P{2.5cm}P{2.5cm}} 
\toprule
\toprule
&{Accuracy}&{AUC}& F-measure&G-means&MCC \\
\midrule
LR&0.4763&0.6885&0.5812&0.6426&0.1967\\
$k$-NN&0.6436&0.7066&0.7298&0.7028&0.2170\\
SVM&0.4900&0.6724& 0.5284&0.6097&0.2813\\
DT&0.6743&0.7028&0.7535&0.7020&0.2172\\
GNB&0.2202&0.5673&0.2703&0.3977&0.0961\\
MLP&\textbf{0.7786}&0.7109&\textbf{0.8276}& 0.7065&0.2537\\
RF&0.7357&0.7525&0.7988&0.7522&0.2823\\
XGB&0.7590&\textbf{0.7595}& 0.8154&\textbf{0.7595}&\textbf{0.2976}\\
ACBR&0.7528&0.7358&0.8106&0.7355&0.2704\\
\bottomrule
\bottomrule
\end{tabular*}
 \begin{tablenotes}
 \normalsize
      \item  \small{\textit{Notes.} The best values of each column are depicted in bold.}
    \end{tablenotes}
  \end{threeparttable}

\end{table}
\subsection{Out-of-sample evaluation with different training metrics}
Table \ref{t:scores_training} shows the out-of-sample prediction results of ACBR system training with different metrics. From the table, we can observe that the ACBR system that utilizes accuracy has the best performance across all criteria. As with other ML models, the empirical findings indicate that accuracy is a reliable evaluation metric for training CBR systems.
\begin{table}[ht!]
    \caption{Out-of-sample prediction results based on different training metrics.}\label{t:scores_training}
 \begin{threeparttable}

 \begin{tabular*}{0.99\textwidth}{p{2cm} P{2.5cm}P{2.5cm}P{2.5cm}P{2.5cm}P{2.5cm}} 
\toprule
\toprule
&{Accuracy}&{AUC}& F-measure&G-means&MCC \\
\midrule
Accuracy&\textbf{0.7528}&\textbf{0.7358}&\textbf{0.8106}&\textbf{0.7355}&\textbf{0.2704}\\
AUC&0.7121&0.7121&0.7814&0.7121&0.2340\\
F-measure&0.7025&0.7106&0.7744&0.7106&0.2303\\
G-means&0.7360&0.7231&0.7986&0.7229& 0.2517\\
MCC& 0.7331 &0.7307&0.7967&0.7308&  0.2589\\

\bottomrule
\bottomrule
\end{tabular*}
 \begin{tablenotes}
 \normalsize
      \item  \small{\textit{Notes.} The best values of each column are depicted in bold.}
    \end{tablenotes}
  \end{threeparttable}

\end{table}
\section{Explainability and decision support}
One of the important characteristics of the CBR method is the interpretability of the prediction results. In this section, we conduct case studies to underline the insights provided by CBR. 
\subsection{Explainability in CBR}
According to \citet{BARREDOARRIETA202082}, the level of explainability of ML models can be evaluated by algorithmic transparency, decomposability and simulatability. Algorithmic transparency is related to the ability of users to understand the process followed by the model to produce any given output from its input data. Decomposability is the ability to interpret individual parts of a model. Simulatability indicates the ability of an ML model to be simulated or thought about strictly by a human. 

The CBR level of explainability is summarized as follows:
\begin{itemize}
\item Algorithmic transparency: It is obvious that analogous problems/cases serve as explanations for human users in the CBR decision-making process \citep{Sormo2005}. However, the similarity measure cannot be fully observed. Thus, some mathematical and statistical analysis will be used for the analysis of the model.

\item Decomposability: The model comprises two functions, i.e., global similarity function and local similarity function. The similarity measure and the set of variables can be decomposed and analyzed separately. However, the number of variables may be too high and/or the similarity measure is too complex. It is not intuitively to understand the model. Thus, visualization will be helpful for user complete comprehension.

\item Simulatability: the complexity of the reasoning process matches human’s native capabilities for simulation and solving new problems. The process is understandable by a human.
\end{itemize}

In this part, we explain and investigate how to employ the CBR system to conduct the analysis of the prediction results and interpret the economic meanings of global and local parameters. 

\subsubsection{Results explanation based on similar cases}
Applying CBR, similar cases can not only be used for prediction but for analogous analysis to detect the reasons for bankruptcy. Assume that there is a query company $Q$ that has been correctly classified as being insolvent by applying the CBR system based on voting from the nine most similar companies (six are insolvent, and three are solvent). To analyze the potential reasons for bankruptcy, the three most similar companies of the company $Q$, $C_1^{d}$ (insolvent), $C_2^d$ (insolvent), and $C_3^n$ (solvent) can be retrieved, and some of their attributes are shown in Table \ref{t:feature_example}. From the table, we can see that company $Q$ obviously has higher sales than the other three companies. Especially, compared with $Q$ and $C_1^{d}$, it is not difficult to find that the company $Q$ adopts a small profits but quick returns operation strategy. Moreover, the profit margin ratios ($\text{profit margin ratio} = {\text{net income}}/{\text{sales}}$) of $Q$, $C_1^{d}$, $C_2^d$ and $C_3^n$ are 2.11\%, 0.87\%, 0.88\%, 5.41\%, respectively. Compared with two insolvent companies $C_1^{d}$ and $C_2^{d}$, the profit margin ratio of the company $Q$ has no obvious increase. The ratio of $Q$ is less than that of $C_3^n$. In addition, $Q$ has a similar amount of accounts receivable with $C_3^n$, while the accounts payable of $Q$ is much higher than that of $C_3^n$. Thus, the fallacious operation strategy would be one of the reasons which led to the bankruptcy of the company $Q$. 
\begin{table}[ht!]
\caption{Features comparison between $Q$ and its three most similar companies, $C_1^{d}$, $C_2^{d}$ and $C_3^{n}$.}\label{t:feature_example}
 \begin{threeparttable}
 \begin{tabular*}{\textwidth}{p{4cm} P{2.5cm} P{2.5cm} P{2.5cm} P{2.5cm}} 
\toprule
\toprule
Feature&${{Q}}$&$C_1^d$&$C_2^d$&${{C}}_3^n$\\
\midrule
Accounts payable      &326,715.51    &243,374.93    &369,152.73    &41,158.13   \\
Accounts receivable   &215,765.17    &205,028.04    &120,153.59    &208,692.67  \\
...     &...    &...  &...    &... \\
{Sales}    &{4,223,270.93}   &{1,614,148.46}   &{1,801,792.58}  & {1,683,185.35}  \\
...     &...    &...  &...    &... \\
{Net income}    &{89,476.07} &{14,316.17}   &{15,850.04}  &{91,298.81} \\
\bottomrule
\bottomrule
\end{tabular*}
  \end{threeparttable}
\end{table}
\subsubsection{Results explanation based on prediction probability}
According to the prediction probability calculation function introduced in the section \ref{Probability}, we can obtain the probability weights of the nine ($K = 9$ in the best-validated CBR) most similar companies $C_1^{d}$, $C_2^{d}$, $C_3^{n}$, $C_4^{n}$, $C_5^{d}$, $C_6^{d}$, $C_7^{d}$, $C_8^{n}$ and $C_9^{d}$, respectively. The prediction probability of $Q$ can be calculated by summing the probability weights of $C_1^{d}$, $C_2^{d}$,  $C_5^{d}$, $C_6^{d}$, $C_7^{d}$ and $C_9^{d}$. Thus, there is a 65.26\% confidence that $Q$ will be an insolvent company.
\subsubsection{Results explanation based on feature relevance}
The weights of the attribute in the CBR global similarity function are important for analyzing the bankruptcy prediction. The weights have been scaled to the relevance scores, the sum of which is equal to 100\%, as shown in Table \ref{t:relevance}. From the table, we can observe that accounts payable is the most important feature, which is money owed by a company to its suppliers shown as a liability on a company's balance sheet. It is reasonable because accounts payable are debts that must be paid off within a given period to avoid default, directly causing insolvency. Besides, we can find that administrative expenses are not important for bankruptcy prediction. It is because the expense commonly incurs that are not directly tied to core functions such as production and sales. 

\begin{table}[t!]
\caption{Feature relevance for insolvency prediction.}\label{t:relevance}
 \begin{threeparttable}
 \begin{tabular*}{\textwidth}{p{5cm}P{2.5cm} p{5cm} P{2.5cm}  } 
\toprule
\toprule
Feature&Relevance (\%) & Feature&Relevance (\%)\\
\midrule
Accounts payable (A.P.) & 6.12 & Interest expenses & 3.80\\
Sales & 5.32 & Bank debt & 3.79\\
Equity & 5.07 & Tangible assets & 3.65\\
Number of employees& 5.06 & Total long-term liabilities & 3.27\\
Total current liabilities & 4.96 & A.R. against affiliated companies & 3.11\\
Operating income & 4.55 & Increase inventories & 3.11\\
Accounts receivable (A.R.) & 4.52 & Increase cash & 2.88\\
Inventories & 4.28 & Increase liabilities & 2.78\\
EBIT & 4.26 & Intangible assets & 2.61\\
Cash & 4.12 & Shareholder loan & 2.59\\
Net income & 3.94 & Accrual for pension liabilities & 2.45\\
Amortization depreciation & 3.91 & A.P. against affiliated companies & 1.20\\
Total assets & 3.82 & Lands & 0.79\\
Current assets & 3.81 & Administrative expenses & 0.24\\
\bottomrule
\bottomrule
\end{tabular*}
 \begin{tablenotes}
 \normalsize
      \item \small{\textit{Notes.} the tables presents the relevance of each feature. The sum of all the value is equal to 100.}
    \end{tablenotes}
  \end{threeparttable}
\end{table}

\begin{figure}[h!t]%
    \centering
    \caption{{Illustration of the local similarity functions of the VAR16 (Sales) based on different parameters $a$ and $b$.}}
    \subfloat[\centering $a_{16}=1,\ b_{16}=1,\ D_{16}=1$]{{\includegraphics[width=0.47\textwidth]{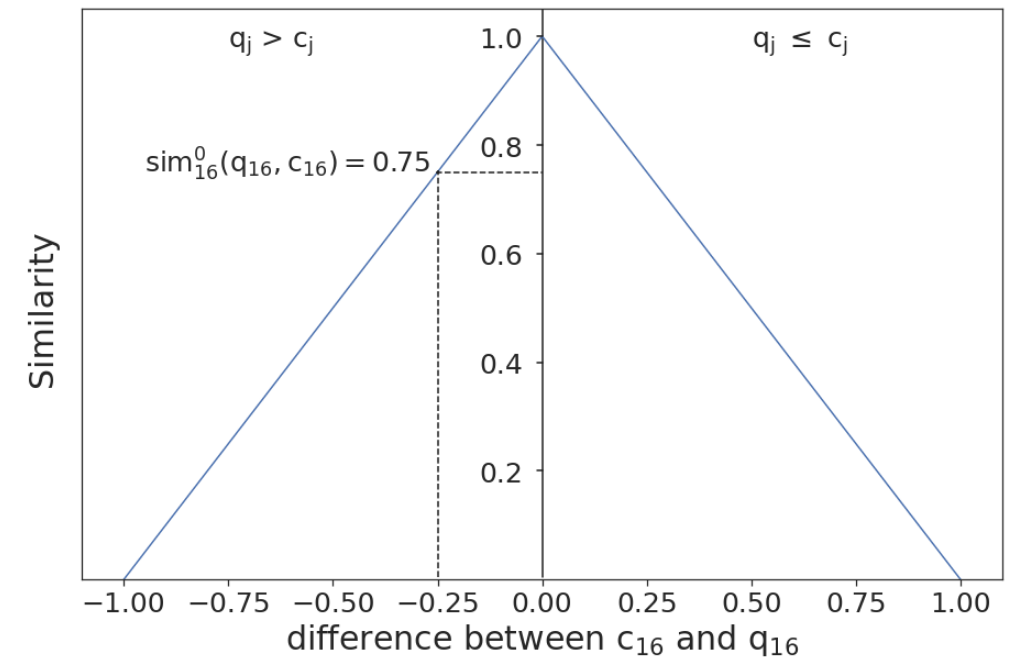} }}%
    \qquad
    \subfloat[\centering $a_{16}=2.12,\ b_{16}=5.53,\ D_{16}=1$]{{\includegraphics[width=0.47\textwidth]{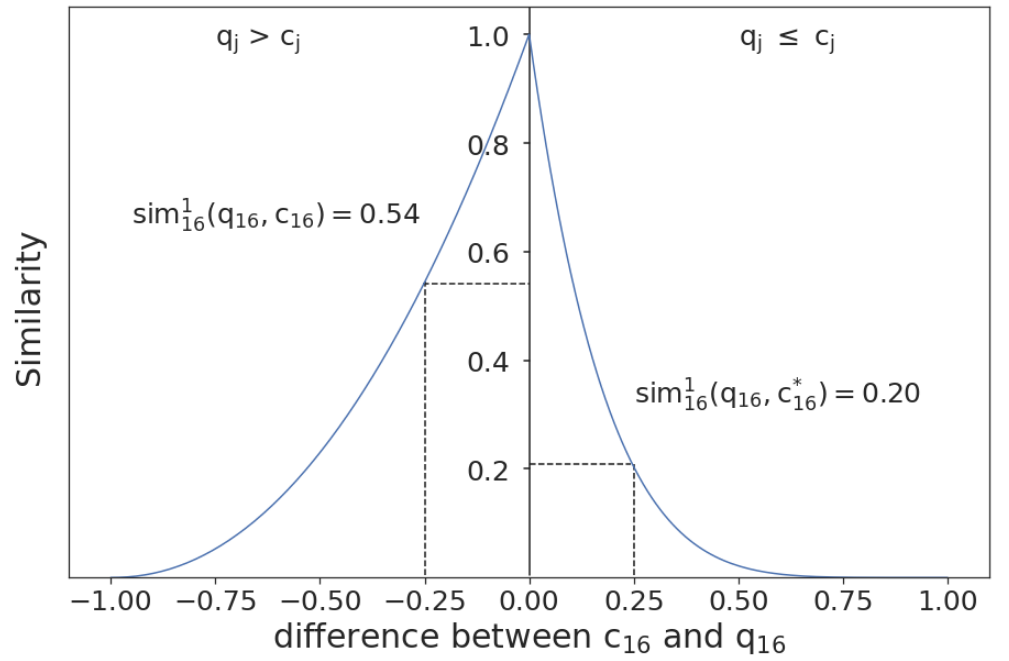} }}%
    \caption*{\small{\textit{Notes}. (a) when $a_{16}=1, b_{16}=1, D_{16}=1$ and $c_{16} - q_{16} = - 0.25$, the similarity $sim^{0}_{16}(q_{16}, c_{16})$ between $c_{16}$ and $q_{16}$ is equal to 0.75. (b) When $a_{16}=2.12, b_{16}=5.53, D_{16}=1$ and $c_{16} - q_{16} = - 0.25$, the similarity $sim^{1}_{16}(q_{16}, c_{16})$ between $c_{16}$ and $q_{16}$ is equal to 0.54. When $a_{16}=2.12, b_{16}=5.53, D_{16}=1$ and $c^{*}_{16} - q_{16} = 0.25$, the similarity $sim^{1}_{16}(q_{16}, c^{*}_{16})$ between $c^{*}_{16}$ and $q_{16}$ is equal to 0.20.}} %
    \label{f:a_b_functions}%
\end{figure}

\begin{centering}
\begin{figure}[ht!]
\caption{Similarities between the query case $q_{16}^m$ and all the $c_{16}$ in dataset based on different parameters $a$ and $b$.}
\centering
\includegraphics[width=0.95\textwidth]{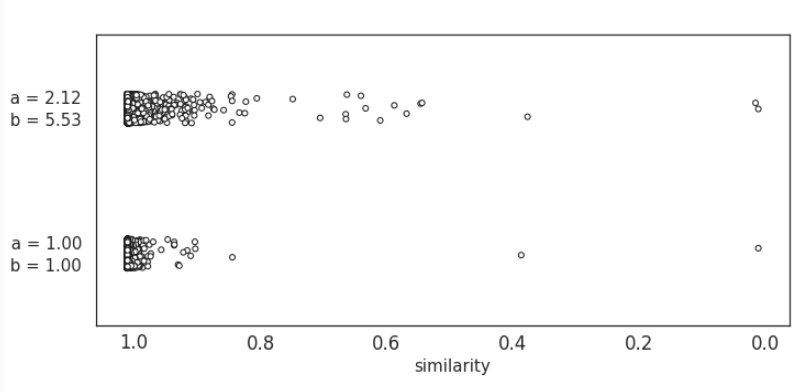}
\caption*{\small{\textit{Notes}. The value of the $q_{16}^m$ is the median value of all the $c_{16}$ in the dataset. }}
\label{f:sales}
\end{figure} 
\end{centering}

In addition, the polynomial characteristics of local functions grant flexibility to pursue more accurate classification. For example,  when calculating the variable similarities between a company $Q$ and a $C_i$ from companies dataset $C$, assume that the difference between the VAR16 (Sales) $c_{16}$ from the $C_i$ and $q_{16}$ from the $Q$ is 0.25, the similarity between the $c_{16}$ and $q_{16}$ can be calculated according to Equation (\ref{eq2}). The illustration of the local similarity functions of the VAR16 is shown in Figure \ref{f:a_b_functions}, in terms of $a_{16}=1$, $b_{16}=1$ (parameters used in EWCBR) and $a_{16}=2.12$, $b_{16}=5.53$ (ACBR), respectively. When $a_{16}=1$ and $b_{16}=1$, the local similarity $\mathrm{sim}^{0}_{16}(q_{16}, c_{16})$ is equal to $0.75$. Meanwhile, when $a_{16}=2.12$ and $b_{16}=
5.53$, the $\mathrm{sim}^{1}_{16}(q_{16}, c_{16})$ is equal to $0.54$. This means that the polynomial function transforms the similarity value between $c_{16}$ and $q_{16}$. Further, the calculated similarities between a $q_{16}^m$ and the $c_{16}$ from all the companies dataset $C$ using the local similarity equations with the two different sets of $a_{16}$ and $b_{16}$ can be seen in Figure \ref{f:sales}. The value of the $q_{16}^m$ is the median value of all the $c_{16}$. From the figure, we can clearly see the relative similarities distance are amplified, which makes companies more distinguishable from one another in terms of the VAR16. Moreover, in Figure \ref{f:a_b_functions}(b), it can be seen that the value of $a_{16}$ is less than $b_{16}$, which means, if the sales absolute difference between a company with smaller sales and a query company is same with the difference between a company with larger sales and the query (e.g. $|c_{16} - q_{16}| = |c^{*}_{16} - q_{16}| = 0.25$, as shown in Figure \ref{f:a_b_functions}(b)), the former is more similar with the query company than the later ($sim^{1}_{16}(q_{16}, c_{16}) = 0.54 > sim^{1}_{16}(q_{16}, c^{*}_{16}) = 0.20$). One reason behind this fact is that a takeoff in sales is difficult to achieve by improving productivity and sales capability, while a decline in sales can occur on the basis of many causes, even a firm operation has not been reorganized \citep{Golder2004207}. Thus, it is reasonable to consider a company with smaller sales to be more similar to a query than the one with larger. 

\begin{figure}

\begin{minipage}[t]{.48\textwidth}
\begin{centering}
  \centering
  \captionof{figure}{Illustration of the local similarity functions of the VAR1 (Cash).}\label{f:cash}
  \includegraphics[width=.9\linewidth]{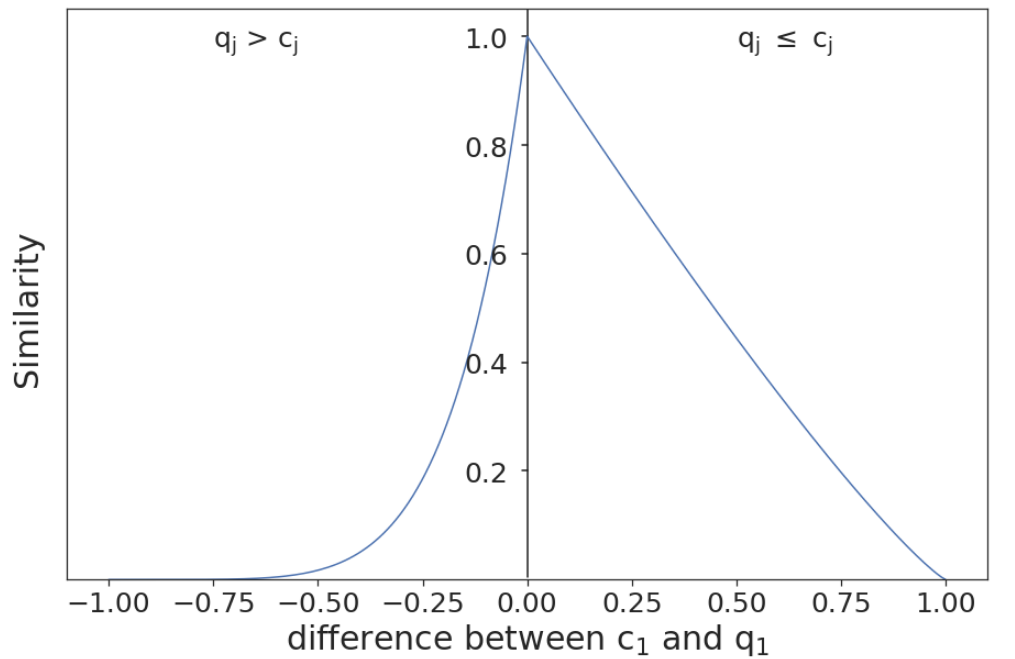}
\end{centering}  
  
  \caption*{\small{\textit{Notes}. The parameters $a_{1}=5.86$ and $b_{1}=1.17$. }}
\end{minipage}%
\hspace{5mm}
\begin{minipage}[t]{.45\textwidth}
  \centering
\captionof{figure}{Illustration of the local similarity functions of the VAR9 (Equity).}\label{f:equity}
  \includegraphics[width=.95\linewidth]{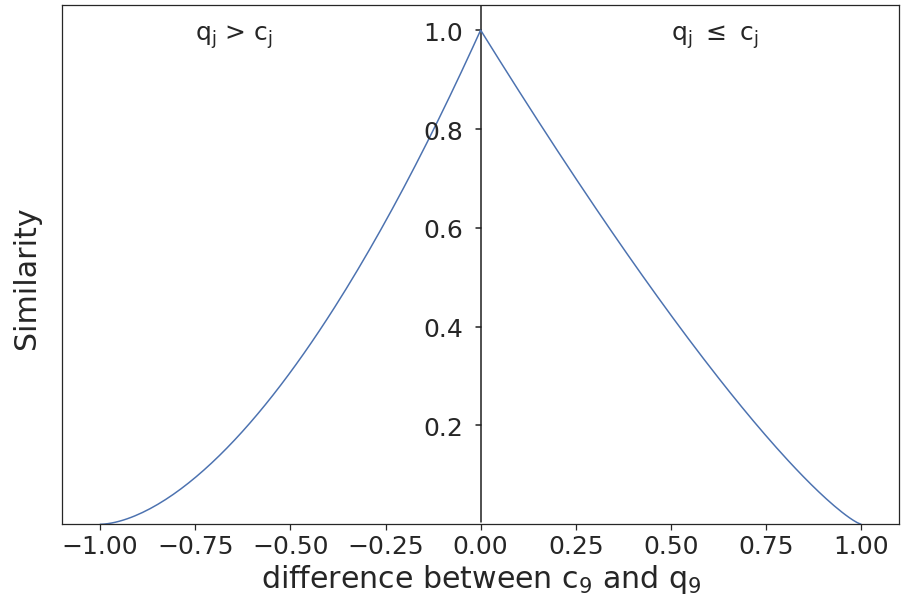}

\caption*{\small{\textit{Notes}. The parameters $a_{9}=1.70$ and $b_{9}=1.24$. }}
\end{minipage}
\end{figure}

Obviously, the varying parameters of the different variables account for their distinct economic significance. For instance, the illustration of the local similarity functions of the VAR1 (Cash) in Figure \ref{f:cash} shows the distinguishing between cash and sales in the similarity calculation. The parameters $a_{1}$ and $b_{1}$ of local similarity are 5.86 and 1.17, respectively. This means that a query company is more likely to be similar to a company holding larger cash. It is reasonable because there is little evidence that excess cash has a substantial short-term effect on capital expenditures, acquisition spending, or dividend payments \citep{OPLER19993}. Meanwhile, operating losses are the leading cause of cash flow constraints in businesses. Consistent with the literature, our findings indicate that the data-driven CBR system can effectively depict the similarity of companies attributed to revealing the nonlinear and asymmetric properties in economic data. It is interesting to find that the relatively stable properties of a company, such as Current assets ($a_{3}=1.95$ and $b_{3}=1.01$), Tangible assets ($a_{4}=2.16$ and $b_{4}=1.38$) and Equity ($a_{9}=1.70$ and $b_{9}=1.24$), did not show an obvious asymmetric local similarity structure. The illustration of the local similarity functions of the VAR9 (Equity) is in Figure \ref{f:equity}. In contrast, intangible assets, which are not physical in nature and stable, such as goodwill, brand recognition, patents, trademarks, and copyrights, present an obvious asymmetric local similarity property ($a_{5}=7.08$ and $b_{5}=2.70$). The findings indicate that elastic features, such as Cash, show an asymmetric similarity structure while static features, such as Equity, show a relatively symmetric similarity structure. 

\begin{figure}[h!]

\begin{minipage}[h!]{.48\textwidth}
  \centering
    \captionof{figure}{Illustration of the local similarity functions of the VAR1 (Cash) and VAR16 (Sales).} \label{f:cash_sales0}
  \includegraphics[width=.95\linewidth]{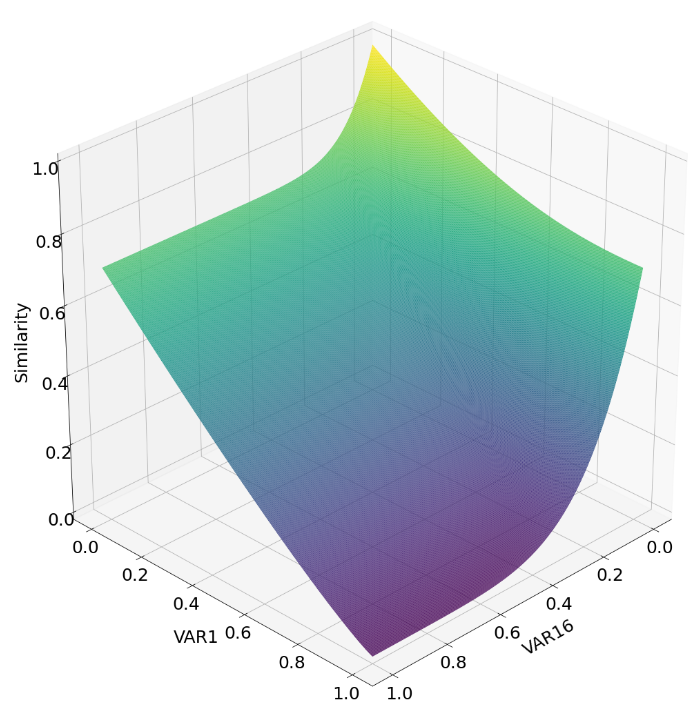}
  \caption*{\small{\textit{Notes}. The parameters $b_{1}=1.17$ and $b_{16}=5.53$, when $q_1\leq c_1$ and $q_{16}\leq c_{16}$}.}
\end{minipage}%
\hspace{5mm}
\begin{minipage}[h!]{.48\textwidth}
  \centering
    \captionof{figure}{Illustration of the local similarity functions of the VAR17 (Administrative expenses) and VAR16 (Sales).} \label{f:adm_sales0}
  \includegraphics[width=.9\linewidth]{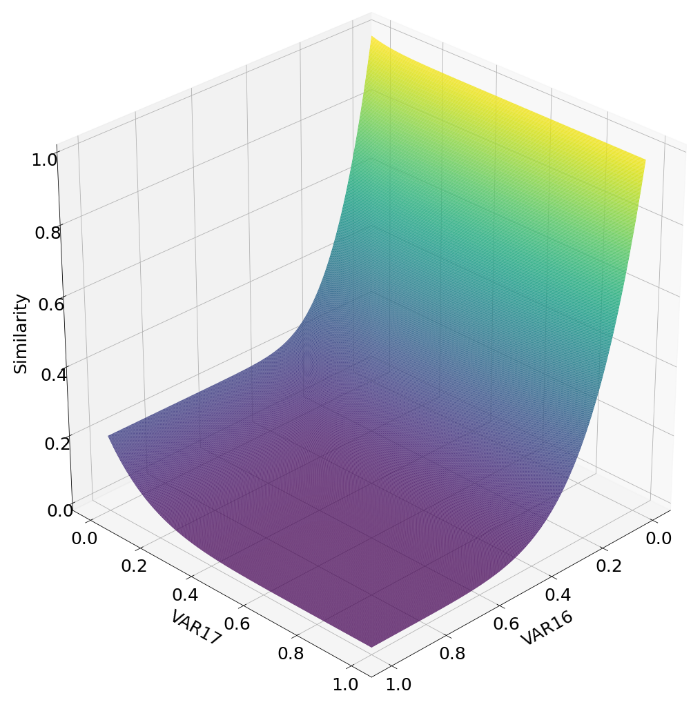}
  \caption*{\small{\textit{Notes}. The parameters $b_{17}=6.53$ and $b_{16}=5.53$, when $q_{17} \leq c_{17}$ and $q_{16}\leq c_{16}$}.}
\end{minipage}
\end{figure}

Further, we investigate the associative effect of features. According to the feature importance in Table \ref{t:relevance}, we assume the weights of Cash (VAR1) and Sales (VAR16) are 0.44 and 0.55, respectively. When comparing a query with a reference company with larger cash and sales ($q_1\leq c_1$ and $q_{16}\leq c_{16}$), the similarity function can be illustrated in Figure \ref{f:cash_sales0}. From the figure, we can observe the asymmetric similarity distribution. In particular, if the query and reference have a similar magnitude of sales ($c_{16} - q_{16}\leq 0.5$), the similarity between them decreases smoothly as the disparity in cash resources increases. Meanwhile, when the query and reference companies do not have comparable sales ($c_{16} - q_{16} > 0.5$), this will lead to more differentiation between firms in the decreasing of the disparity in cash resources. In addition, the weights of features, obviously, have a significant impact on their contribution to the computation of similarity. As shown in Figure \ref{f:adm_sales0}, Administrative expenses (VAR17) have a limited influence on weighing up similarity in comparison to Sales. The rest parameters $a$ and $b$ of all the variables are shown in Appendix G.

\subsection{Decision support for bankruptcy}
In this section, we explore the application of CBR prediction probability to provide decision support regarding bankruptcy by conducting the case study of the prediction and analysis for an insolvent German company, i.e., Gerry Weber International AG. In particular, we first explain the results of Shapley-CBR. Then, through the case study of Gerry Weber, we show the application of CBR prediction probability as a bankruptcy risk indicator and elaborate on the decision support based on CBR system for the insolvent company. 
\subsubsection{Shapley-CBR}
\begin{centering}
\begin{figure}[ht!]
\caption{Average absolute impact on solvent prediction confidence based on Shapley-CBR.}
\centering
\includegraphics[width=\textwidth]{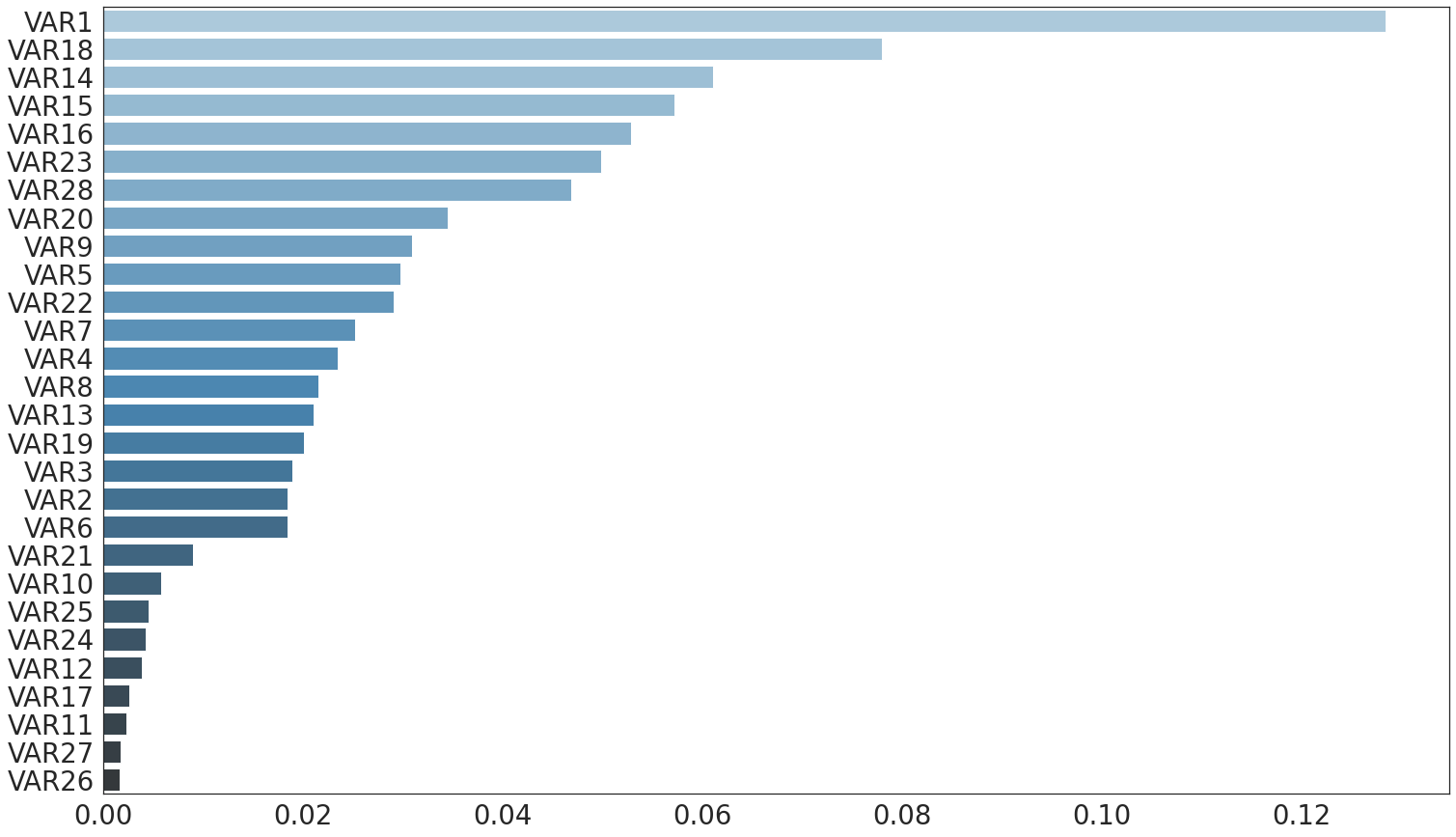}
\label{f:shapley}
\end{figure} 
\end{centering}

As aforementioned, the Shapley value can be used to explain the contributions of variables to the improvement of solvent confidence. The average absolute impact of Shapley value on the confidence is shown in Figure \ref{f:shapley}. From the figure, the results indicate that VAR1 (Cash) has the most decisive influence. It is reasonable because cash on hand is the most liquid type of asset. The most direct way to solve insolvency via investment is cash injection. An interesting point to note from the results is the importance of the variable VAR18 (Amortization and Depreciation) ranks second. Amortisation is the process by which debt is paid off with a fixed repayment schedule in regular installments over a certain time period. The insolvent company can negotiate with its creditors and modify the amortization profile of their benefits. For instance, the loan repayments being recalculated on a longer amortization profile can help the company resolve insolvency; meanwhile, it will give creditors get a better chance of recovering their debts compared with the situation the company goes into liquidation. Establishing effective reorganization proceedings with creditors is significantly helpful for the company to get through the financial distress. Moreover, the third and fourth important variables are VAR14 (Bank debt) and VAR15 (Account payable). Similar to amortization, negotiation with banks and vendors or suppliers to delay or restructure the debts or goods payments is an efficient way to increase the probability of discontinuing insolvency proceedings. By contrast, from Figure \ref{f:shapley}, we can find that administrative expenses (VAR12), accrual for pension liabilities (VAR17), A.P. against affiliated companies (VAR27), A.R. against affiliated companies (VAR26) are not important, which is consistent with our intuition. 

\subsubsection{Case study with Gerry Weber}
Gerry Weber is a German fashion manufacturer and retailer based in Halle, established in 1973. In recent years, the fashion empire suffered from the decline in customer footfall in the city centers and the triumph of online trading. The self-administered insolvency proceedings opened in 2019. To get through the financial distress, several efforts have been adapted: 1) More than a hundred branches were closed, and numerous jobs were cut; 2) The existing shareholders were forced out of the company without compensation; 3) The financial investors are invited, which provide up to 49 million euros to satisfy its creditors and to finance operations. The insolvency proceedings claim was withdrawn in 2020. From Table \ref{t:gw}, we can see that, in 2019, Gerry Weber filed for insolvency as predicted, based on the financial status in the fiscal year 2018. In the following year, after experiencing active self-remedy, the firm was solvent as predicted, and the confidence score recovered to 76\%.     
 
\begin{figure}[h!]
\centering
\caption{Data plot of Gerry Weber and its most similar companies from 2017 to 2019 in terms of VAR9, VAR15 and VAR21. The blue and red dots stand for solvent and insolvent companies, respectively. The green dot stands for Gerry Weber. }\label{f:changes_weber}%
\subfloat[\centering{Gerry Weber in 2017}\label{f:weber2017}]{{\includegraphics[width=7.5cm]{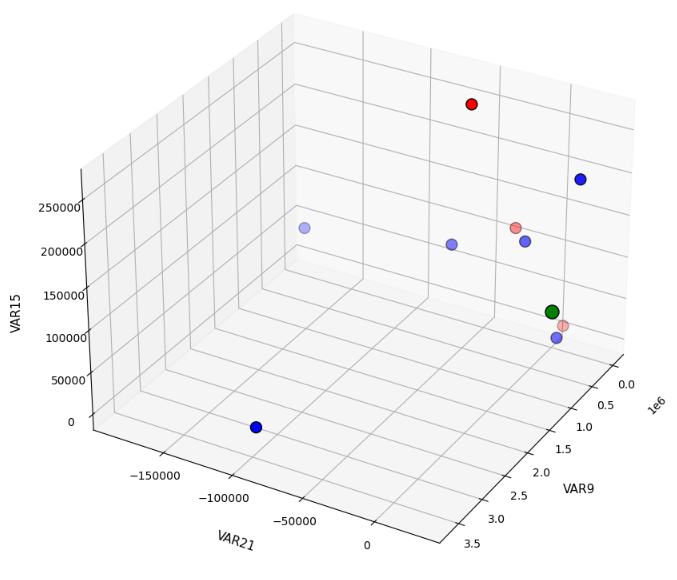} }}%
\qquad
\subfloat[\centering{Gerry Weber in 2018}\label{f:weber2018}]{{\includegraphics[width=7.5cm]{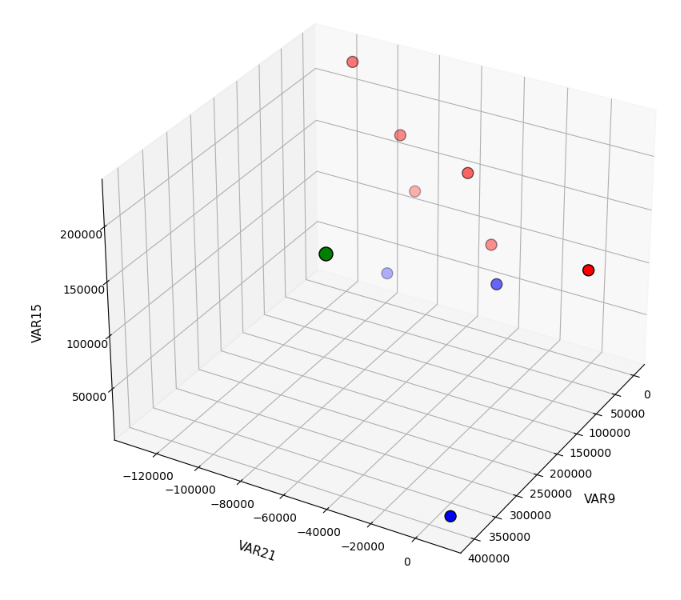} }}%
\qquad
\subfloat[\centering{Gerry Weber in 2019}\label{f:weber2019}]{{\includegraphics[width=7.5cm]{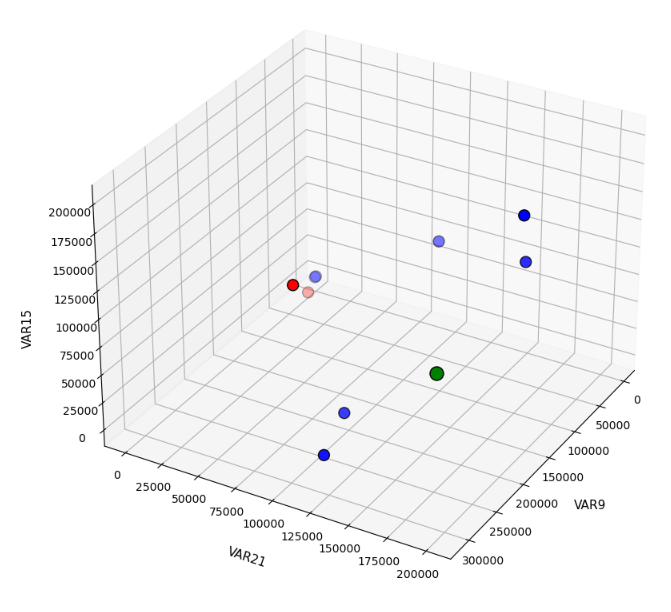} }}%

\end{figure}

\begin{table}[ht!]
\caption{Insolvent prediction on Gerry Weber and the prediction confidence.}\label{t:gw}
 \begin{threeparttable}
 \begin{tabular*}{\textwidth}{p{2cm}p{2cm} P{2cm} P{9cm}} 
\toprule
\toprule
Year&Prediction&Confidence&Description\\
\midrule
2017&Solvent &74\%& \multirow{5}{9cm}{Gerry Weber filed for insolvency in January 2019. \\ Approximately 120 shops in Germany and in addition about 180 sales points in Europe were closed in April 2019. \\The insolvency proceedings were discontinued in January 2020.}\\
2018&Solvent &74\% &\\
2019&\textbf{Insolvent} &\textbf{65\%} &\\
2020&Solvent &76\%&\\
2021&Solvent &62\%& \\
\bottomrule
\bottomrule
\end{tabular*}
 \begin{tablenotes}
 \scriptsize
    \end{tablenotes}
  \end{threeparttable}
\end{table}
According to the financial variables from 2017 to 2019, which can be found in Appendix H, it is not surprising to find that Equity (VAR9) and Operation income (VAR21) experienced the most significant changes since it is intuitive to increase income and assets while decreasing the liabilities to solve insolvency. Figure \ref{f:changes_weber} shows the alteration of its most similar reference companies in terms of the two variables against the other important feature, Accounts payable (VAR15). The green dot stands for Gerry Weber, while the blue and red dots stand for solvent and insolvent companies, respectively. From Figure \ref{f:changes_weber}(a), we can observe that the majority of the most similar companies have negative operation income in comparison to Gerry Weber. This indicates that the operating income of Gerry Weber tends to be negative. In the fiscal year 2018, as aforementioned, Gerry Weber was predicted to be insolvent in the following year. From Figure \ref{f:changes_weber}(b), we can observe that the operating income of Gerry Weber became negative and was the worst among its most similar referencing firms. It is interesting to find that all insolvent companies have larger accounts payable than Gerry Weber, and solvent similar companies have an identical level of account payable, which indicates that the accounts payable is not the main reason behind the insolvency of Gerry Weber. From Figure \ref{f:changes_weber}(c), we can find that Gerry Weber was dedicated to increasing VAR21 and VAR9, contributing to the discontinuance of the insolvency proceedings.

\begin{centering}
\begin{figure}[ht!]
\caption{The dynamics of the prediction probability as the accumulation of the feature variable changes.}
\centering
\includegraphics[width=\textwidth]{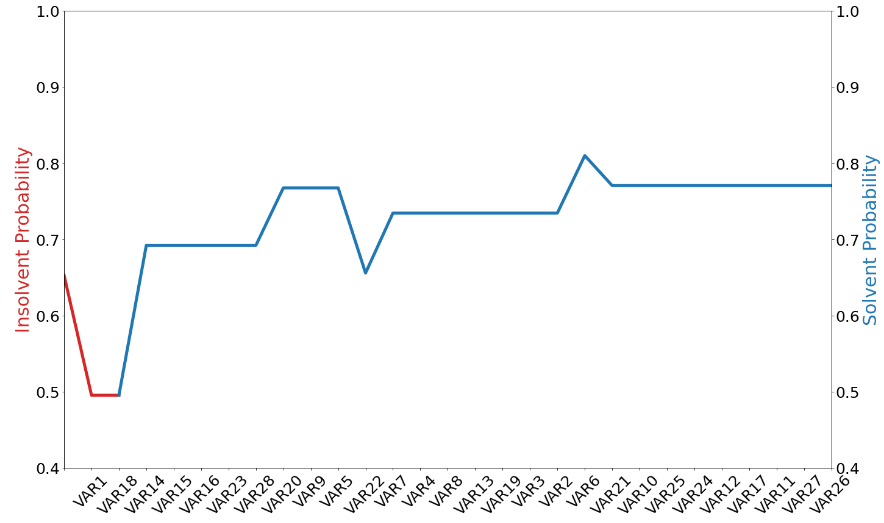}
\label{f:prob_changes}
\end{figure} 
\end{centering}

\begin{figure}[h!]
\centering
\caption{Data plot of Gerry Weber and its most similar companies, without changing VAR22 and with changing VAR22, in terms of VAR9, VAR15
and VAR21. The blue and red dots stand for solvent and insolvent companies, respectively. The green dot stands for Gerry Weber.}\label{f:changes_VAR22}%
\subfloat[\centering{Without changing VAR22}]{{\includegraphics[width=7.5cm]{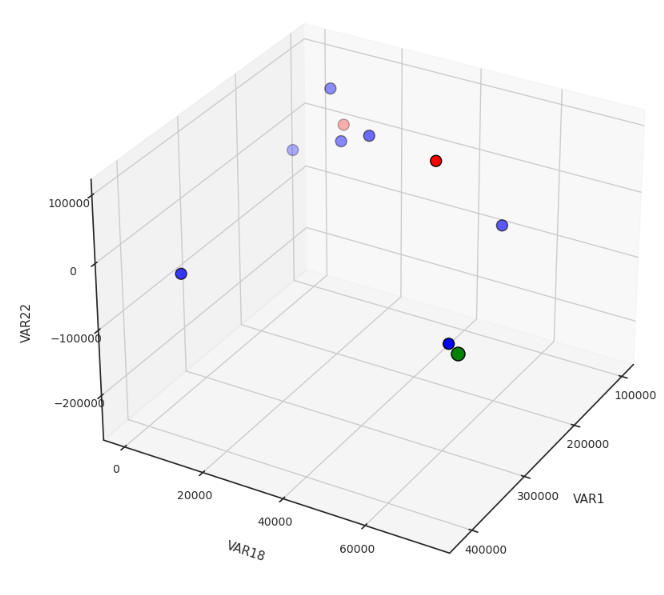} }}%
\qquad
\subfloat[\centering{With changing VAR22}]{{\includegraphics[width=7.5cm]{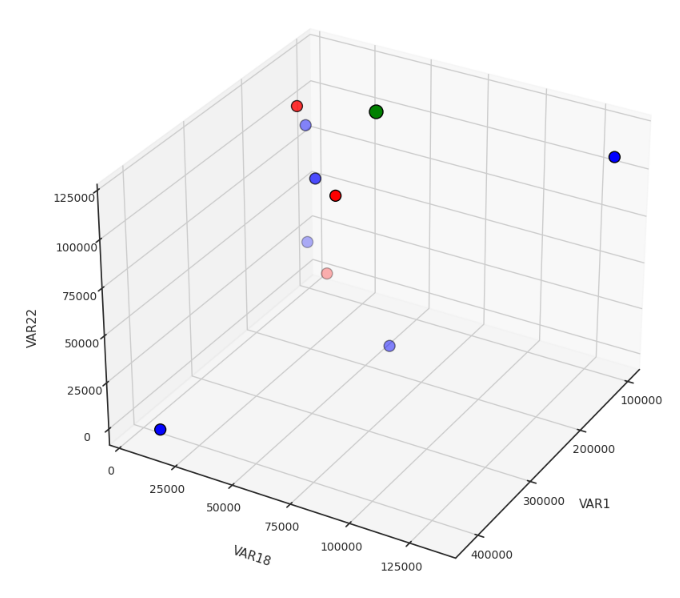} }}%

\end{figure}

In addition, based on the analysis of the Shapley-CBR, we know the ranking of feature variables to increase the solvent prediction probability. However, from the three figures in Figure \ref{f:changes_weber}, we can observe that the company try to reduce the accounts payable, one of the most important features influencing prediction decision and prediction probability, but the changes are not significant. Thus, it is also important to investigate the efficiency of the applied solutions to solve the insolvency problem. In particular, we accumulatively replaced the feature variables of Gerry Weber in 2018 with the ones in 2019 following the feature ranking of the Shapley-CBR. From Figure \ref{f:prob_changes}, we can find that the change of VAR1 has reduced the insolvent probability. The accumulation of VAR1, VAR18 and VAR14 has reversed the insolvent situation, giving a 69\% prediction probability for solvency. Subsequently, the rest of the changes are accumulating the solvency prediction confidence. It is interesting to find that the changes in VAR22 (Operating income) and VAR21 (Net income) decreased the solvency prediction confidence. In order to investigate the impact of the change of the VAR22, the data plot of Gerry Weber and its most similar reference companies in terms of the VAR22 against VAR1 and VAR18 is shown in Figure \ref{f:changes_VAR22}. From the figure, we can find that the increase in operating income makes Gerry Weber be more close to insolvent companies. The findings indicate that the increase in operating income and net income is not an efficient way to get rid of insolvency. The company in the financial dilemma should put more endeavour into changing the other financial status. For instance, we know that Operating Income = Gross Profit - Operating Expenses - Depreciation - Amortization. As we discussed, the reduction of depreciation and amortization or operating expenses, such as the deduction of staff wages or the layoff of employees, can effectively contribute to solvency. However, the extremely pursuing of such reduction targeting an increase in operating income does not benefit solving the insolvency problem.

\section{Conclusion}
This paper investigates a data-driven CBR system for bankruptcy evaluation. The traditional CBR highly depends on domain knowledge and prior experience, and the manually-design suffers from a low prediction accuracy. We apply an automatic procedure to design the retrieving in the CBR system. The parameters in the similarity calculation functions are automatically optimized, targeting the increase in prediction accuracy. The experimental results show that the CBR method performs competitively compared to the widely used classification ML methods. Moreover, the CBR approach distinguishes itself from other black-box ML methods by showing the interpretability of the prediction process and further providing useful information for bankruptcy-related decision-making. This characteristic is helpful for companies' stakeholders, decision makers, and investors. In particular, we examine the level of the explainability of the CBR system in detail by analyzing the bankruptcy prediction process and conducting experiments to outline a unified view of explanation in the CBR system. The results show the scheme to explain the decision-making process based on the CBR system and explore more decisive information. Further, we extensively examine the efficiency of the local similarity functions in CBR system. The empirical results reveal the importance of asymmetrical polynomial local similarity function to increase the prediction accuracy in distance/similarity-based classification and exhibit the advantages how the asymmetrical polynomial parameters can better reflect practical economic meanings in the comparison of similar companies. The latter enhances the users' conference and provides useful information to support stakeholders in making decisions. Our findings demonstrate a promising direction for future research by establishing a well-designed explainable model that can improve not only prediction accuracy but also increase explainability.

For future studies, several extensions of the current study can be developed. In the proposed approach, the design of global similarity is based on the existing feature scoring methods. In further research, we will explore a general way to detect the optimal feature weights. In addition, the maintenance of a CBR system is necessary, the costs of which may include the cost of adding new cases to strengthen the case base and refining existing cases \citep{Bensoussan2009294}. Maximization of the system effectiveness in supporting bankruptcy-related decisions will be worth investigating. Besides, the current study was carried out using the bankruptcy dataset, but the generality of the proposed CBR system ensures a possible application to other decision-support systems.

\newpage
\bibliographystyle{model2-names}\biboptions{authoryear}
\bibliography{mybibfile}

\begin{thebibliography}{77}
\expandafter\ifx\csname natexlab\endcsname\relax\def\natexlab#1{#1}\fi
\providecommand{\url}[1]{\texttt{#1}}
\providecommand{\href}[2]{#2}
\providecommand{\path}[1]{#1}
\providecommand{\DOIprefix}{doi:}
\providecommand{\ArXivprefix}{arXiv:}
\providecommand{\URLprefix}{URL: }
\providecommand{\Pubmedprefix}{pmid:}
\providecommand{\doi}[1]{\href{http://dx.doi.org/#1}{\path{#1}}}
\providecommand{\Pubmed}[1]{\href{pmid:#1}{\path{#1}}}
\providecommand{\bibinfo}[2]{#2}
\ifx\xfnm\relax \def\xfnm[#1]{\unskip,\space#1}\fi
\bibitem[{Aamodt and Plaza(1994)}]{Aamodt1994}
\bibinfo{author}{Aamodt, A}, \bibinfo{author}{Plaza, E} (\bibinfo{year}{1994}).
\newblock \bibinfo{title}{Case-based reasoning: Foundational issues,
  methodological variations, and system approaches}.
\newblock \textit{\bibinfo{journal}{AI Communications}} \bibinfo{volume}{7}:
  \bibinfo{pages}{39--59}.
\newblock \DOIprefix\doi{10.3233/AIC-1994-7104}. \bibinfo{note}{1}.
\bibitem[{Ahn and Kim(2009)}]{AHN2009599}
\bibinfo{author}{Ahn, H}, \bibinfo{author}{Kim, K} (\bibinfo{year}{2009}).
\newblock \bibinfo{title}{Bankruptcy prediction modeling with hybrid case-based
  reasoning and genetic algorithms approach}.
\newblock \textit{\bibinfo{journal}{Applied Soft Computing}}
  \bibinfo{volume}{9}(\bibinfo{number}{2}): \bibinfo{pages}{599--607}.
\newblock \DOIprefix\doi{10.1016/j.asoc.2008.08.002}.
\bibitem[{Altman(1968)}]{Edward1968589}
\bibinfo{author}{Altman, EI} (\bibinfo{year}{1968}).
\newblock \bibinfo{title}{Financial ratios, discriminant analysis and the
  prediction of corporate bankruptcy}.
\newblock \textit{\bibinfo{journal}{The Journal of Finance}}
  \bibinfo{volume}{23}(\bibinfo{number}{4}): \bibinfo{pages}{589--609}.
\newblock \URLprefix \url{http://www.jstor.org/stable/2978933}.
\bibitem[{Bach and Althoff(2012)}]{Bach201217}
\bibinfo{author}{Bach, K}, \bibinfo{author}{Althoff, KD}
  (\bibinfo{year}{2012}).
\newblock \bibinfo{title}{Developing case-based reasoning applications using
  mycbr 3}, in: \bibinfo{editor}{Agudo, BD}, \bibinfo{editor}{Watson, I}
  (Eds.), \bibinfo{booktitle}{Case-Based Reasoning Research and Development},
  \textit{\bibinfo{publisher}{Springer Berlin Heidelberg},
  \bibinfo{address}{Berlin, Heidelberg}}.  \bibinfo{pages}{17--31}.
\bibitem[{Balog et~al.(2017)Balog, Bátyi, Csóka and Pintér}]{BALOG2017614}
\bibinfo{author}{Balog, D}, \bibinfo{author}{Bátyi, TL},
  \bibinfo{author}{Csóka, P}, \bibinfo{author}{Pintér, M}
  (\bibinfo{year}{2017}).
\newblock \bibinfo{title}{Properties and comparison of risk capital allocation
  methods}.
\newblock \textit{\bibinfo{journal}{European Journal of Operational Research}}
  \bibinfo{volume}{259}(\bibinfo{number}{2}): \bibinfo{pages}{614--625}.
\newblock \DOIprefix\doi{https://doi.org/10.1016/j.ejor.2016.10.052}.
\bibitem[{{Barredo Arrieta} et~al.(2020){Barredo Arrieta}, Díaz-Rodríguez,
  {Del Ser}, Bennetot, Tabik, Barbado, Garcia, Gil-Lopez, Molina, Benjamins,
  Chatila and Herrera}]{BARREDOARRIETA202082}
\bibinfo{author}{{Barredo Arrieta}, A}, \bibinfo{author}{Díaz-Rodríguez, N},
  \bibinfo{author}{{Del Ser}, J}, \bibinfo{author}{Bennetot, A},
  \bibinfo{author}{Tabik, S}, \bibinfo{author}{Barbado, A},
  \bibinfo{author}{Garcia, S}, \bibinfo{author}{Gil-Lopez, S},
  \bibinfo{author}{Molina, D}, \bibinfo{author}{Benjamins, R},
  \bibinfo{author}{Chatila, R}, \bibinfo{author}{Herrera, F}
  (\bibinfo{year}{2020}).
\newblock \bibinfo{title}{Explainable artificial intelligence (xai): Concepts,
  taxonomies, opportunities and challenges toward responsible ai}.
\newblock \textit{\bibinfo{journal}{Information Fusion}} \bibinfo{volume}{58}:
  \bibinfo{pages}{82--115}.
\newblock \DOIprefix\doi{https://doi.org/10.1016/j.inffus.2019.12.012}.
\bibitem[{Beddoe and Petrovic(2006)}]{Beddoe_EJOR}
\bibinfo{author}{Beddoe, GR}, \bibinfo{author}{Petrovic, S}
  (\bibinfo{year}{2006}).
\newblock \bibinfo{title}{Selecting and weighting features using a genetic
  algorithm in a case-based reasoning approach to personnel rostering}.
\newblock \textit{\bibinfo{journal}{European Journal of Operational Research}}
  \bibinfo{volume}{175}(\bibinfo{number}{2}): \bibinfo{pages}{649--671}.
\newblock \DOIprefix\doi{https://doi.org/10.1016/j.ejor.2004.12.028}.
\bibitem[{Bensoussan et~al.(2009)Bensoussan, Mookerjee, Mookerjee and
  Yue}]{Bensoussan2009294}
\bibinfo{author}{Bensoussan, A}, \bibinfo{author}{Mookerjee, R},
  \bibinfo{author}{Mookerjee, V}, \bibinfo{author}{Yue, WT}
  (\bibinfo{year}{2009}).
\newblock \bibinfo{title}{Maintaining diagnostic knowledge-based systems: A
  control-theoretic approach}.
\newblock \textit{\bibinfo{journal}{Management Science}}
  \bibinfo{volume}{55}(\bibinfo{number}{2}): \bibinfo{pages}{294--310}.
\newblock \DOIprefix\doi{10.1287/mnsc.1080.0908}.
\bibitem[{Bent{\'e}jac et~al.(2021)Bent{\'e}jac, Cs{\"o}rg{\H{o}} and
  Mart{\'i}nez-Mu{\~{n}}oz}]{Bentejac2021}
\bibinfo{author}{Bent{\'e}jac, C}, \bibinfo{author}{Cs{\"o}rg{\H{o}}, A},
  \bibinfo{author}{Mart{\'i}nez-Mu{\~{n}}oz, G} (\bibinfo{year}{2021}).
\newblock \bibinfo{title}{A comparative analysis of gradient boosting
  algorithms}.
\newblock \textit{\bibinfo{journal}{Artificial Intelligence Review}}
  \bibinfo{volume}{54}(\bibinfo{number}{3}): \bibinfo{pages}{1937--1967}.
\newblock \DOIprefix\doi{10.1007/s10462-020-09896-5}.
\bibitem[{Berganti{\~{n}}os and Lorenzo-Freire(2008)}]{Bergantinos2008}
\bibinfo{author}{Berganti{\~{n}}os, G}, \bibinfo{author}{Lorenzo-Freire, S}
  (\bibinfo{year}{2008}).
\newblock \bibinfo{title}{``optimistic'' weighted shapley rules in minimum cost
  spanning tree problems}.
\newblock \textit{\bibinfo{journal}{European Journal of Operational Research}}
  \bibinfo{volume}{185}(\bibinfo{number}{1}): \bibinfo{pages}{289--298}.
\bibitem[{Bernanke(1981)}]{Bernanke1981}
\bibinfo{author}{Bernanke, BS} (\bibinfo{year}{1981}).
\newblock \bibinfo{title}{Bankruptcy, liquidity, and recession}.
\newblock \textit{\bibinfo{journal}{The American Economic Review}}
  \bibinfo{volume}{71}(\bibinfo{number}{2}): \bibinfo{pages}{155--159}.
\newblock \URLprefix \url{http://www.jstor.org/stable/1815710}.
\bibitem[{Bradley(1997)}]{BRADLEY19971145}
\bibinfo{author}{Bradley, AP} (\bibinfo{year}{1997}).
\newblock \bibinfo{title}{The use of the area under the roc curve in the
  evaluation of machine learning algorithms}.
\newblock \textit{\bibinfo{journal}{Pattern Recognition}}
  \bibinfo{volume}{30}(\bibinfo{number}{7}): \bibinfo{pages}{1145--1159}.
\newblock \DOIprefix\doi{10.1016/S0031-3203(96)00142-2}.
\bibitem[{Branco et~al.(2016)Branco, Torgo and Ribeiro}]{Branco2016}
\bibinfo{author}{Branco, P}, \bibinfo{author}{Torgo, L},
  \bibinfo{author}{Ribeiro, RP} (\bibinfo{year}{2016}).
\newblock \bibinfo{title}{A survey of predictive modeling on imbalanced
  domains}.
\newblock \textit{\bibinfo{journal}{ACM Computing Surveys}}
  \bibinfo{volume}{49}(\bibinfo{number}{2}).
\newblock \DOIprefix\doi{10.1145/2907070}.
\bibitem[{Breiman(2001)}]{Breiman2001}
\bibinfo{author}{Breiman, L} (\bibinfo{year}{2001}).
\newblock \bibinfo{title}{Random forests}.
\newblock \textit{\bibinfo{journal}{Machine Learning}}
  \bibinfo{volume}{45}(\bibinfo{number}{1}): \bibinfo{pages}{5--32}.
\newblock \DOIprefix\doi{10.1023/A:1010933404324}.
\bibitem[{Cappelen et~al.(2019)Cappelen, Luttens, S{\o}rensen and
  Tungodden}]{Cappelen20192832}
\bibinfo{author}{Cappelen, AW}, \bibinfo{author}{Luttens, RI},
  \bibinfo{author}{S{\o}rensen, E{\O}}, \bibinfo{author}{Tungodden, B}
  (\bibinfo{year}{2019}).
\newblock \bibinfo{title}{Fairness in bankruptcies: An experimental study}.
\newblock \textit{\bibinfo{journal}{Management Science}}
  \bibinfo{volume}{65}(\bibinfo{number}{6}): \bibinfo{pages}{2832--2841}.
\newblock \DOIprefix\doi{10.1287/mnsc.2018.3029}.
\bibitem[{Ceriani and Verme(2012)}]{Ceriani2012}
\bibinfo{author}{Ceriani, L}, \bibinfo{author}{Verme, P}
  (\bibinfo{year}{2012}).
\newblock \bibinfo{title}{The origins of the gini index: extracts from
  variabilit{\`a} e mutabilit{\`a} (1912) by corrado gini}.
\newblock \textit{\bibinfo{journal}{The Journal of Economic Inequality}}
  \bibinfo{volume}{10}(\bibinfo{number}{3}): \bibinfo{pages}{421--443}.
\newblock \DOIprefix\doi{10.1007/s10888-011-9188-x}.
\bibitem[{Chi et~al.(1993)Chi, Chen and Kiang}]{CHI199367}
\bibinfo{author}{Chi, RT}, \bibinfo{author}{Chen, M}, \bibinfo{author}{Kiang,
  MY} (\bibinfo{year}{1993}).
\newblock \bibinfo{title}{Generalized case-based reasoning system for portfolio
  management}.
\newblock \textit{\bibinfo{journal}{Expert Systems with Applications}}
  \bibinfo{volume}{6}(\bibinfo{number}{1}): \bibinfo{pages}{67 -- 76}.
\newblock \DOIprefix\doi{10.1016/0957-4174(93)90019-3}. \bibinfo{note}{special
  Issue: Case-Based Reasoning and its Applications}.
\bibitem[{Chuang(2013)}]{CHUANG2013174}
\bibinfo{author}{Chuang, CL} (\bibinfo{year}{2013}).
\newblock \bibinfo{title}{Application of hybrid case-based reasoning for
  enhanced performance in bankruptcy prediction}.
\newblock \textit{\bibinfo{journal}{Information Sciences}}
  \bibinfo{volume}{236}: \bibinfo{pages}{174--185}.
\newblock \DOIprefix\doi{10.1016/j.ins.2013.02.015}.
\bibitem[{Cost and Salzberg(1993)}]{Cost1993}
\bibinfo{author}{Cost, S}, \bibinfo{author}{Salzberg, S}
  (\bibinfo{year}{1993}).
\newblock \bibinfo{title}{A weighted nearest neighbor algorithm for learning
  with symbolic features}.
\newblock \textit{\bibinfo{journal}{Machine Learning}}
  \bibinfo{volume}{10}(\bibinfo{number}{1}): \bibinfo{pages}{57--78}.
\newblock \DOIprefix\doi{10.1023/A:1022664626993}.
\bibitem[{Csóka et~al.(2022)Csóka, Illés and Solymosi}]{CSOKA2022378}
\bibinfo{author}{Csóka, P}, \bibinfo{author}{Illés, F},
  \bibinfo{author}{Solymosi, T} (\bibinfo{year}{2022}).
\newblock \bibinfo{title}{On the shapley value of liability games}.
\newblock \textit{\bibinfo{journal}{European Journal of Operational Research}}
  \bibinfo{volume}{300}(\bibinfo{number}{1}): \bibinfo{pages}{378--386}.
\newblock \DOIprefix\doi{https://doi.org/10.1016/j.ejor.2021.10.012}.
\bibitem[{Cui et~al.(2006)Cui, Wong and Lui}]{Cui2006597}
\bibinfo{author}{Cui, G}, \bibinfo{author}{Wong, ML}, \bibinfo{author}{Lui, HK}
  (\bibinfo{year}{2006}).
\newblock \bibinfo{title}{Machine learning for direct marketing response
  models: Bayesian networks with evolutionary programming}.
\newblock \textit{\bibinfo{journal}{Management Science}}
  \bibinfo{volume}{52}(\bibinfo{number}{4}): \bibinfo{pages}{597--612}.
\newblock \DOIprefix\doi{10.1287/mnsc.1060.0514}.
\bibitem[{Cunningham et~al.(2003)Cunningham, Doyle and
  Loughrey}]{Cunningham2003122}
\bibinfo{author}{Cunningham, P}, \bibinfo{author}{Doyle, D},
  \bibinfo{author}{Loughrey, J} (\bibinfo{year}{2003}).
\newblock \bibinfo{title}{An evaluation of the usefulness of case-based
  explanation}, in: \bibinfo{editor}{Ashley, KD}, \bibinfo{editor}{Bridge, DG}
  (Eds.), \bibinfo{booktitle}{Case-Based Reasoning Research and Development},
  \textit{\bibinfo{publisher}{Springer Berlin Heidelberg},
  \bibinfo{address}{Berlin, Heidelberg}}.  \bibinfo{pages}{122--130}.
\bibitem[{Dastile et~al.(2020)Dastile, Celik and Potsane}]{DASTILE2020106263}
\bibinfo{author}{Dastile, X}, \bibinfo{author}{Celik, T},
  \bibinfo{author}{Potsane, M} (\bibinfo{year}{2020}).
\newblock \bibinfo{title}{Statistical and machine learning models in credit
  scoring: A systematic literature survey}.
\newblock \textit{\bibinfo{journal}{Applied Soft Computing}}
  \bibinfo{volume}{91}: \bibinfo{pages}{106263}.
\newblock \DOIprefix\doi{https://doi.org/10.1016/j.asoc.2020.106263}.
\bibitem[{Deng(1996)}]{Deng_EJOR}
\bibinfo{author}{Deng, PS} (\bibinfo{year}{1996}).
\newblock \bibinfo{title}{Using case-based reasoning approach to the support of
  ill-structured decisions}.
\newblock \textit{\bibinfo{journal}{European Journal of Operational Research}}
  \bibinfo{volume}{93}(\bibinfo{number}{3}): \bibinfo{pages}{511--521}.
\newblock \DOIprefix\doi{https://doi.org/10.1016/0377-2217(95)00081-X}.
\bibitem[{Ding et~al.(2012)Ding, Tian, Yu and Guo}]{Ding2012990}
\bibinfo{author}{Ding, AA}, \bibinfo{author}{Tian, S}, \bibinfo{author}{Yu, Y},
  \bibinfo{author}{Guo, H} (\bibinfo{year}{2012}).
\newblock \bibinfo{title}{A class of discrete transformation survival models
  with application to default probability prediction}.
\newblock \textit{\bibinfo{journal}{Journal of the American Statistical
  Association}} \bibinfo{volume}{107}(\bibinfo{number}{499}):
  \bibinfo{pages}{990--1003}.
\newblock \DOIprefix\doi{10.1080/01621459.2012.682806}.
\bibitem[{Dustmann et~al.(2014)Dustmann, Fitzenberger, Schönberg and
  Spitz-Oener}]{Dustmann2014167}
\bibinfo{author}{Dustmann, C}, \bibinfo{author}{Fitzenberger, B},
  \bibinfo{author}{Schönberg, U}, \bibinfo{author}{Spitz-Oener, A}
  (\bibinfo{year}{2014}).
\newblock \bibinfo{title}{From sick man of europe to economic superstar:
  Germany's resurgent economy}.
\newblock \textit{\bibinfo{journal}{Journal of Economic Perspectives}}
  \bibinfo{volume}{28}(\bibinfo{number}{1}): \bibinfo{pages}{167--88}.
\newblock \DOIprefix\doi{10.1257/jep.28.1.167}.
\bibitem[{Gardner and Dorling(1998)}]{GARDNER19982627}
\bibinfo{author}{Gardner, M}, \bibinfo{author}{Dorling, S}
  (\bibinfo{year}{1998}).
\newblock \bibinfo{title}{Artificial neural networks (the multilayer
  perceptron)—a review of applications in the atmospheric sciences}.
\newblock \textit{\bibinfo{journal}{Atmospheric Environment}}
  \bibinfo{volume}{32}(\bibinfo{number}{14}): \bibinfo{pages}{2627--2636}.
\newblock \DOIprefix\doi{10.1016/S1352-2310(97)00447-0}.
\bibitem[{Golder and Tellis(2004)}]{Golder2004207}
\bibinfo{author}{Golder, PN}, \bibinfo{author}{Tellis, GJ}
  (\bibinfo{year}{2004}).
\newblock \bibinfo{title}{Growing, growing, gone: Cascades, diffusion, and
  turning points in the product life cycle}.
\newblock \textit{\bibinfo{journal}{Marketing Science}}
  \bibinfo{volume}{23}(\bibinfo{number}{2}): \bibinfo{pages}{207--218}.
\newblock \DOIprefix\doi{10.1287/mksc.1040.0057}.
\bibitem[{Guessoum et~al.(2014)Guessoum, Laskri and Lieber}]{GUESSOUM2014267}
\bibinfo{author}{Guessoum, S}, \bibinfo{author}{Laskri, MT},
  \bibinfo{author}{Lieber, J} (\bibinfo{year}{2014}).
\newblock \bibinfo{title}{Respidiag: A case-based reasoning system for the
  diagnosis of chronic obstructive pulmonary disease}.
\newblock \textit{\bibinfo{journal}{Expert Systems with Applications}}
  \bibinfo{volume}{41}(\bibinfo{number}{2}): \bibinfo{pages}{267--273}.
\newblock \DOIprefix\doi{10.1016/j.eswa.2013.05.065}.
\bibitem[{H{\"a}rdle et~al.(2011)H{\"a}rdle, Hoffmann and Moro}]{Hardle2011}
\bibinfo{author}{H{\"a}rdle, WK}, \bibinfo{author}{Hoffmann, L},
  \bibinfo{author}{Moro, R} (\bibinfo{year}{2011}).
\newblock \bibinfo{title}{Learning machines supporting bankruptcy prediction}.
  \textit{\bibinfo{publisher}{Springer Berlin Heidelberg},
  \bibinfo{address}{Berlin, Heidelberg}}.
\newblock  \bibinfo{pages}{225--250}.
\newblock \DOIprefix\doi{10.1007/978-3-642-18062-0_7}.
\bibitem[{H{\"a}rdle et~al.(2005)H{\"a}rdle, Moro and Sch{\"a}fer}]{Hardle2005}
\bibinfo{author}{H{\"a}rdle, WK}, \bibinfo{author}{Moro, R},
  \bibinfo{author}{Sch{\"a}fer, D} (\bibinfo{year}{2005}).
\newblock \bibinfo{title}{Predicting Bankruptcy with Support Vector Machines}.
  \textit{\bibinfo{publisher}{Springer Berlin Heidelberg},
  \bibinfo{address}{Berlin, Heidelberg}}.
\newblock  \bibinfo{pages}{225--248}.
\newblock \DOIprefix\doi{10.1007/3-540-27395-6_10}.
\bibitem[{Hart(2017)}]{Hart2017}
\bibinfo{author}{Hart, S} (\bibinfo{year}{2017}).
\newblock \bibinfo{title}{Shapley Value}. \textit{\bibinfo{publisher}{Palgrave
  Macmillan UK}, \bibinfo{address}{London}}.
\newblock  \bibinfo{pages}{1--5}.
\newblock \DOIprefix\doi{10.1057/978-1-349-95121-5_1369-2}.
\bibitem[{Hosaka(2019)}]{HOSAKA2019287}
\bibinfo{author}{Hosaka, T} (\bibinfo{year}{2019}).
\newblock \bibinfo{title}{Bankruptcy prediction using imaged financial ratios
  and convolutional neural networks}.
\newblock \textit{\bibinfo{journal}{Expert Systems with Applications}}
  \bibinfo{volume}{117}: \bibinfo{pages}{287--299}.
\newblock \DOIprefix\doi{10.1016/j.eswa.2018.09.039}.
\bibitem[{Ince(2014)}]{INCE2014205}
\bibinfo{author}{Ince, H} (\bibinfo{year}{2014}).
\newblock \bibinfo{title}{Short term stock selection with case-based reasoning
  technique}.
\newblock \textit{\bibinfo{journal}{Applied Soft Computing}}
  \bibinfo{volume}{22}: \bibinfo{pages}{205--212}.
\newblock \DOIprefix\doi{10.1016/j.asoc.2014.05.017}.
\bibitem[{Jabeur et~al.(2021)Jabeur, Mefteh-Wali and Viviani}]{Jabeur2021}
\bibinfo{author}{Jabeur, SB}, \bibinfo{author}{Mefteh-Wali, S},
  \bibinfo{author}{Viviani, JL} (\bibinfo{year}{2021}).
\newblock \bibinfo{title}{Forecasting gold price with the xgboost algorithm and
  shap interaction values}.
\newblock \textit{\bibinfo{journal}{Annals of Operations Research}}
  \DOIprefix\doi{10.1007/s10479-021-04187-w}.
\bibitem[{Jardin(2021)}]{DUJARDIN2021869}
\bibinfo{author}{Jardin, PD} (\bibinfo{year}{2021}).
\newblock \bibinfo{title}{Forecasting corporate failure using ensemble of
  self-organizing neural networks}.
\newblock \textit{\bibinfo{journal}{European Journal of Operational Research}}
  \bibinfo{volume}{288}(\bibinfo{number}{3}): \bibinfo{pages}{869--885}.
\newblock \DOIprefix\doi{10.1016/j.ejor.2020.06.020}.
\bibitem[{Jo et~al.(1997)Jo, Han and Lee}]{JO199797}
\bibinfo{author}{Jo, H}, \bibinfo{author}{Han, I}, \bibinfo{author}{Lee, H}
  (\bibinfo{year}{1997}).
\newblock \bibinfo{title}{Bankruptcy prediction using case-based reasoning,
  neural networks, and discriminant analysis}.
\newblock \textit{\bibinfo{journal}{Expert Systems with Applications}}
  \bibinfo{volume}{13}(\bibinfo{number}{2}): \bibinfo{pages}{97--108}.
\newblock \DOIprefix\doi{10.1016/S0957-4174(97)00011-0}.
\bibitem[{Kleinbaum(1994)}]{Kleinbaum1994}
\bibinfo{author}{Kleinbaum, DG} (\bibinfo{year}{1994}).
\newblock \bibinfo{title}{Introduction to Logistic Regression}.
  \textit{\bibinfo{publisher}{Springer New York}, \bibinfo{address}{New York,
  NY}}.
\newblock  \bibinfo{pages}{1--38}.
\newblock \DOIprefix\doi{10.1007/978-1-4757-4108-7_1}.
\bibitem[{Kononenko et~al.(1997)Kononenko, {\v{S}}imec and
  Robnik-{\v{S}}ikonja}]{Kononenko1997}
\bibinfo{author}{Kononenko, I}, \bibinfo{author}{{\v{S}}imec, E},
  \bibinfo{author}{Robnik-{\v{S}}ikonja, M} (\bibinfo{year}{1997}).
\newblock \bibinfo{title}{Overcoming the myopia of inductive learning
  algorithms with relieff}.
\newblock \textit{\bibinfo{journal}{Applied Intelligence}}
  \bibinfo{volume}{7}(\bibinfo{number}{1}): \bibinfo{pages}{39--55}.
\newblock \DOIprefix\doi{10.1023/A:1008280620621}.
\bibitem[{Kraskov et~al.(2004)Kraskov, St\"ogbauer and
  Grassberger}]{Kraskov2004}
\bibinfo{author}{Kraskov, A}, \bibinfo{author}{St\"ogbauer, H},
  \bibinfo{author}{Grassberger, P} (\bibinfo{year}{2004}).
\newblock \bibinfo{title}{Estimating mutual information}.
\newblock \textit{\bibinfo{journal}{Physical Review E}} \bibinfo{volume}{69}:
  \bibinfo{pages}{066138}.
\newblock \DOIprefix\doi{10.1103/PhysRevE.69.066138}.
\bibitem[{Kullback(1959)}]{Kullback59}
\bibinfo{author}{Kullback, S} (\bibinfo{year}{1959}).
\newblock \textit{\bibinfo{title}{Information Theory and Statistics}}.
\newblock \bibinfo{publisher}{Wiley}, \bibinfo{address}{New York}.
\bibitem[{Lamy et~al.(2019)Lamy, Sekar, Guezennec, Bouaud and
  Séroussi}]{LAMY201942}
\bibinfo{author}{Lamy, JB}, \bibinfo{author}{Sekar, B},
  \bibinfo{author}{Guezennec, G}, \bibinfo{author}{Bouaud, J},
  \bibinfo{author}{Séroussi, B} (\bibinfo{year}{2019}).
\newblock \bibinfo{title}{Explainable artificial intelligence for breast
  cancer: A visual case-based reasoning approach}.
\newblock \textit{\bibinfo{journal}{Artificial Intelligence in Medicine}}
  \bibinfo{volume}{94}: \bibinfo{pages}{42--53}.
\newblock \DOIprefix\doi{10.1016/j.artmed.2019.01.001}.
\bibitem[{Lensberg et~al.(2006)Lensberg, Eilifsen and McKee}]{LENSBERG2006677}
\bibinfo{author}{Lensberg, T}, \bibinfo{author}{Eilifsen, A},
  \bibinfo{author}{McKee, TE} (\bibinfo{year}{2006}).
\newblock \bibinfo{title}{Bankruptcy theory development and classification via
  genetic programming}.
\newblock \textit{\bibinfo{journal}{European Journal of Operational Research}}
  \bibinfo{volume}{169}(\bibinfo{number}{2}): \bibinfo{pages}{677--697}.
\newblock \DOIprefix\doi{10.1016/j.ejor.2004.06.013}. \bibinfo{note}{feature
  Cluster on Scatter Search Methods for Optimization}.
\bibitem[{Li and Sun(2009)}]{LI200910085}
\bibinfo{author}{Li, H}, \bibinfo{author}{Sun, J} (\bibinfo{year}{2009}).
\newblock \bibinfo{title}{Predicting business failure using multiple case-based
  reasoning combined with support vector machine}.
\newblock \textit{\bibinfo{journal}{Expert Systems with Applications}}
  \bibinfo{volume}{36}(\bibinfo{number}{6}): \bibinfo{pages}{10085--10096}.
\newblock \DOIprefix\doi{https://doi.org/10.1016/j.eswa.2009.01.013}.
\bibitem[{Li et~al.(2018)Li, Jamieson, DeSalvo, Rostamizadeh and
  Talwalkar}]{li20181}
\bibinfo{author}{Li, L}, \bibinfo{author}{Jamieson, K},
  \bibinfo{author}{DeSalvo, G}, \bibinfo{author}{Rostamizadeh, A},
  \bibinfo{author}{Talwalkar, A} (\bibinfo{year}{2018}).
\newblock \bibinfo{title}{Hyperband: A novel bandit-based approach to
  hyperparameter optimization}.
\newblock \textit{\bibinfo{journal}{Journal of Machine Learning Research}}
  \bibinfo{volume}{18}(\bibinfo{number}{185}): \bibinfo{pages}{1--52}.
\newblock \URLprefix \url{http://jmlr.org/papers/v18/16-558.html}.
\bibitem[{Li and Becker(2021)}]{Li2021}
\bibinfo{author}{Li, W}, \bibinfo{author}{Becker, DM} (\bibinfo{year}{2021}).
\newblock \bibinfo{title}{Day-ahead electricity price prediction applying
  hybrid models of lstm-based deep learning methods and feature selection
  algorithms under consideration of market coupling}.
\newblock \textit{\bibinfo{journal}{Energy}} \bibinfo{volume}{237}:
  \bibinfo{pages}{121543}.
\bibitem[{Li et~al.(2022)Li, Paraschiv and Sermpinis}]{Li20221}
\bibinfo{author}{Li, W}, \bibinfo{author}{Paraschiv, F},
  \bibinfo{author}{Sermpinis, G} (\bibinfo{year}{2022}).
\newblock \bibinfo{title}{A data-driven explainable case-based reasoning
  approach for financial risk detection}.
\newblock \textit{\bibinfo{journal}{Quantitative Finance}}
  \bibinfo{volume}{0}(\bibinfo{number}{0}): \bibinfo{pages}{1--18}.
\newblock \DOIprefix\doi{10.1080/14697688.2022.2118071}.
\bibitem[{Lin and Ding(2011)}]{LIN201164}
\bibinfo{author}{Lin, H}, \bibinfo{author}{Ding, H} (\bibinfo{year}{2011}).
\newblock \bibinfo{title}{Predicting ion channels and their types by the
  dipeptide mode of pseudo amino acid composition}.
\newblock \textit{\bibinfo{journal}{Journal of Theoretical Biology}}
  \bibinfo{volume}{269}(\bibinfo{number}{1}): \bibinfo{pages}{64 -- 69}.
\newblock \DOIprefix\doi{10.1016/j.jtbi.2010.10.019}.
\bibitem[{Lundberg et~al.(2020)Lundberg, Erion, Chen, DeGrave, Prutkin, Nair,
  Katz, Himmelfarb, Bansal and Lee}]{Lundberg2020}
\bibinfo{author}{Lundberg, SM}, \bibinfo{author}{Erion, G},
  \bibinfo{author}{Chen, H}, \bibinfo{author}{DeGrave, A},
  \bibinfo{author}{Prutkin, JM}, \bibinfo{author}{Nair, B},
  \bibinfo{author}{Katz, R}, \bibinfo{author}{Himmelfarb, J},
  \bibinfo{author}{Bansal, N}, \bibinfo{author}{Lee, SI}
  (\bibinfo{year}{2020}).
\newblock \bibinfo{title}{From local explanations to global understanding with
  explainable ai for trees}.
\newblock \textit{\bibinfo{journal}{Nature Machine Intelligence}}
  \bibinfo{volume}{2}(\bibinfo{number}{1}): \bibinfo{pages}{56--67}.
\newblock \DOIprefix\doi{10.1038/s42256-019-0138-9}.
\bibitem[{Lundberg and Lee(2017)}]{Lundberg2017}
\bibinfo{author}{Lundberg, SM}, \bibinfo{author}{Lee, SI}
  (\bibinfo{year}{2017}).
\newblock \bibinfo{title}{A unified approach to interpreting model
  predictions}, in: \bibinfo{booktitle}{Proceedings of the 31st International
  Conference on Neural Information Processing Systems},
  \textit{\bibinfo{publisher}{Curran Associates Inc.}, \bibinfo{address}{Red
  Hook, NY, USA}}. p. \bibinfo{pages}{4768–4777}.
\bibitem[{Löw et~al.(2019)Löw, Hesser and Blessing}]{LOW2019103127}
\bibinfo{author}{Löw, N}, \bibinfo{author}{Hesser, J},
  \bibinfo{author}{Blessing, M} (\bibinfo{year}{2019}).
\newblock \bibinfo{title}{Multiple retrieval case-based reasoning for
  incomplete datasets}.
\newblock \textit{\bibinfo{journal}{Journal of Biomedical Informatics}}
  \bibinfo{volume}{92}: \bibinfo{pages}{103127}.
\newblock \DOIprefix\doi{10.1016/j.jbi.2019.103127}.
\bibitem[{Mai et~al.(2019)Mai, Tian, Lee and Ma}]{MAI2019743}
\bibinfo{author}{Mai, F}, \bibinfo{author}{Tian, S}, \bibinfo{author}{Lee, C},
  \bibinfo{author}{Ma, L} (\bibinfo{year}{2019}).
\newblock \bibinfo{title}{Deep learning models for bankruptcy prediction using
  textual disclosures}.
\newblock \textit{\bibinfo{journal}{European Journal of Operational Research}}
  \bibinfo{volume}{274}(\bibinfo{number}{2}): \bibinfo{pages}{743--758}.
\newblock \DOIprefix\doi{10.1016/j.ejor.2018.10.024}.
\bibitem[{Martin(1977)}]{MARTIN1977249}
\bibinfo{author}{Martin, D} (\bibinfo{year}{1977}).
\newblock \bibinfo{title}{Early warning of bank failure: A logit regression
  approach}.
\newblock \textit{\bibinfo{journal}{Journal of Banking \& Finance}}
  \bibinfo{volume}{1}(\bibinfo{number}{3}): \bibinfo{pages}{249--276}.
\newblock \DOIprefix\doi{10.1016/0378-4266(77)90022-X}.
\bibitem[{Mitchell(1997)}]{Mitchell1997}
\bibinfo{author}{Mitchell, TM} (\bibinfo{year}{1997}).
\newblock \textit{\bibinfo{title}{Machine Learning}}.
\newblock \bibinfo{edition}{1} ed., \bibinfo{publisher}{McGraw-Hill, Inc.},
  \bibinfo{address}{USA}.
\bibitem[{Morris(1994)}]{Morris1994}
\bibinfo{author}{Morris, BW} (\bibinfo{year}{1994}).
\newblock \bibinfo{title}{Scan: A case-based reasoning model for generating
  information system control recommendations}.
\newblock \textit{\bibinfo{journal}{Intelligent Systems in Accounting, Finance
  and Management}} \bibinfo{volume}{3}(\bibinfo{number}{1}):
  \bibinfo{pages}{47--63}.
\newblock \DOIprefix\doi{10.1002/j.1099-1174.1994.tb00054.x}.
\bibitem[{Moxey et~al.(2010)Moxey, Robertson, Newby, Hains, Williamson and
  Pearson}]{Moxey201025}
\bibinfo{author}{Moxey, A}, \bibinfo{author}{Robertson, J},
  \bibinfo{author}{Newby, D}, \bibinfo{author}{Hains, I},
  \bibinfo{author}{Williamson, M}, \bibinfo{author}{Pearson, SA}
  (\bibinfo{year}{2010}).
\newblock \bibinfo{title}{{Computerized clinical decision support for
  prescribing: provision does not guarantee uptake}}.
\newblock \textit{\bibinfo{journal}{Journal of the American Medical Informatics
  Association}} \bibinfo{volume}{17}(\bibinfo{number}{1}):
  \bibinfo{pages}{25--33}.
\newblock \DOIprefix\doi{10.1197/jamia.M3170}.
\bibitem[{Nagel and Purnanandam(2019)}]{Nagel20192421}
\bibinfo{author}{Nagel, S}, \bibinfo{author}{Purnanandam, A}
  (\bibinfo{year}{2019}).
\newblock \bibinfo{title}{Banks' risk dynamics and distance to default}.
\newblock \textit{\bibinfo{journal}{The Review of Financial Studies}}
  \bibinfo{volume}{33}(\bibinfo{number}{6}): \bibinfo{pages}{2421--2467}.
\newblock \DOIprefix\doi{10.1093/rfs/hhz125}.
\bibitem[{Nishihara and Shibata(2021)}]{NISHIHARA20211017}
\bibinfo{author}{Nishihara, M}, \bibinfo{author}{Shibata, T}
  (\bibinfo{year}{2021}).
\newblock \bibinfo{title}{The effects of asset liquidity on dynamic sell-out
  and bankruptcy decisions}.
\newblock \textit{\bibinfo{journal}{European Journal of Operational Research}}
  \bibinfo{volume}{288}(\bibinfo{number}{3}): \bibinfo{pages}{1017--1035}.
\newblock \DOIprefix\doi{10.1016/j.ejor.2020.06.031}.
\bibitem[{Novaković(2011)}]{Novakovic2016}
\bibinfo{author}{Novaković, J} (\bibinfo{year}{2011}).
\newblock \bibinfo{title}{Toward optimal feature selection using ranking
  methods and classification algorithms}.
\newblock \textit{\bibinfo{journal}{Yugoslav Journal of Operations Research}}
  \bibinfo{volume}{21}(\bibinfo{number}{1}).
\newblock \DOIprefix\doi{0.2298/YJOR1101119N}.
\bibitem[{Opler et~al.(1999)Opler, Pinkowitz, Stulz and
  Williamson}]{OPLER19993}
\bibinfo{author}{Opler, T}, \bibinfo{author}{Pinkowitz, L},
  \bibinfo{author}{Stulz, R}, \bibinfo{author}{Williamson, R}
  (\bibinfo{year}{1999}).
\newblock \bibinfo{title}{The determinants and implications of corporate cash
  holdings}.
\newblock \textit{\bibinfo{journal}{Journal of Financial Economics}}
  \bibinfo{volume}{52}(\bibinfo{number}{1}): \bibinfo{pages}{3--46}.
\newblock \DOIprefix\doi{https://doi.org/10.1016/S0304-405X(99)00003-3}.
\bibitem[{O'Roarty et~al.(1997)O'Roarty, Patterson, McGreal and
  Adair}]{OROARTY1997417}
\bibinfo{author}{O'Roarty, B}, \bibinfo{author}{Patterson, D},
  \bibinfo{author}{McGreal, S}, \bibinfo{author}{Adair, A}
  (\bibinfo{year}{1997}).
\newblock \bibinfo{title}{A case-based reasoning approach to the selection of
  comparable evidence for retail rent determination}.
\newblock \textit{\bibinfo{journal}{Expert Systems with Applications}}
  \bibinfo{volume}{12}(\bibinfo{number}{4}): \bibinfo{pages}{417 -- 428}.
\newblock \DOIprefix\doi{10.1016/S0957-4174(97)83769-4}.
\bibitem[{Richter and Weber(2013)}]{Richter2013}
\bibinfo{author}{Richter, MM}, \bibinfo{author}{Weber, RO}
  (\bibinfo{year}{2013}).
\newblock \textit{\bibinfo{title}{Case-Based Reasoning: A Textbook}}.
\newblock \bibinfo{publisher}{Springer Publishing Company, Incorporated}.
\bibitem[{Rudin(2019)}]{Rudin2019}
\bibinfo{author}{Rudin, C} (\bibinfo{year}{2019}).
\newblock \bibinfo{title}{Stop explaining black box machine learning models for
  high stakes decisions and use interpretable models instead}.
\newblock \textit{\bibinfo{journal}{Nature Machine Intelligence}}
  \bibinfo{volume}{1}(\bibinfo{number}{5}): \bibinfo{pages}{206--215}.
\newblock \DOIprefix\doi{10.1038/s42256-019-0048-x}.
\bibitem[{Shin and Lee(2002)}]{SHIN2002321}
\bibinfo{author}{Shin, KS}, \bibinfo{author}{Lee, YJ} (\bibinfo{year}{2002}).
\newblock \bibinfo{title}{A genetic algorithm application in bankruptcy
  prediction modeling}.
\newblock \textit{\bibinfo{journal}{Expert Systems with Applications}}
  \bibinfo{volume}{23}(\bibinfo{number}{3}): \bibinfo{pages}{321--328}.
\newblock \DOIprefix\doi{10.1016/S0957-4174(02)00051-9}.
\bibitem[{Song and Lu(2015)}]{Song2015}
\bibinfo{author}{Song, YY}, \bibinfo{author}{Lu, Y} (\bibinfo{year}{2015}).
\newblock \bibinfo{title}{Decision tree methods: applications for
  classification and prediction}.
\newblock \textit{\bibinfo{journal}{Shanghai Archives of Psychiatry}}
  \bibinfo{volume}{27}(\bibinfo{number}{2}): \bibinfo{pages}{130--135}.
\newblock \URLprefix \url{https://pubmed.ncbi.nlm.nih.gov/26120265}.
\bibitem[{S{\o}rmo et~al.(2005)S{\o}rmo, Cassens and Aamodt}]{Sormo2005}
\bibinfo{author}{S{\o}rmo, F}, \bibinfo{author}{Cassens, J},
  \bibinfo{author}{Aamodt, A} (\bibinfo{year}{2005}).
\newblock \bibinfo{title}{Explanation in case-based reasoning--perspectives and
  goals}.
\newblock \textit{\bibinfo{journal}{Artificial Intelligence Review}}
  \bibinfo{volume}{24}(\bibinfo{number}{2}): \bibinfo{pages}{109--143}.
\newblock \DOIprefix\doi{10.1007/s10462-005-4607-7}.
\bibitem[{Stoltzfus(2011)}]{Stoltzfus20111099}
\bibinfo{author}{Stoltzfus, JC} (\bibinfo{year}{2011}).
\newblock \bibinfo{title}{Logistic regression: A brief primer}.
\newblock \textit{\bibinfo{journal}{Academic Emergency Medicine}}
  \bibinfo{volume}{18}(\bibinfo{number}{10}): \bibinfo{pages}{1099--1104}.
\newblock \DOIprefix\doi{10.1111/j.1553-2712.2011.01185.x}.
\bibitem[{Sun and Shenoy(2007)}]{SUN2007738}
\bibinfo{author}{Sun, L}, \bibinfo{author}{Shenoy, PP} (\bibinfo{year}{2007}).
\newblock \bibinfo{title}{Using bayesian networks for bankruptcy prediction:
  Some methodological issues}.
\newblock \textit{\bibinfo{journal}{European Journal of Operational Research}}
  \bibinfo{volume}{180}(\bibinfo{number}{2}): \bibinfo{pages}{738--753}.
\newblock \DOIprefix\doi{10.1016/j.ejor.2006.04.019}.
\bibitem[{Tian and Yu(2017)}]{TIAN2017510}
\bibinfo{author}{Tian, S}, \bibinfo{author}{Yu, Y} (\bibinfo{year}{2017}).
\newblock \bibinfo{title}{Financial ratios and bankruptcy predictions: An
  international evidence}.
\newblock \textit{\bibinfo{journal}{International Review of Economics \&
  Finance}} \bibinfo{volume}{51}: \bibinfo{pages}{510--526}.
\newblock \DOIprefix\doi{10.1016/j.iref.2017.07.025}.
\bibitem[{Tian et~al.(2015)Tian, Yu and Guo}]{TIAN201589}
\bibinfo{author}{Tian, S}, \bibinfo{author}{Yu, Y}, \bibinfo{author}{Guo, H}
  (\bibinfo{year}{2015}).
\newblock \bibinfo{title}{Variable selection and corporate bankruptcy
  forecasts}.
\newblock \textit{\bibinfo{journal}{Journal of Banking \& Finance}}
  \bibinfo{volume}{52}: \bibinfo{pages}{89--100}.
\newblock \DOIprefix\doi{10.1016/j.jbankfin.2014.12.003}.
\bibitem[{Vapnik(1998)}]{Vapnik1998}
\bibinfo{author}{Vapnik, VN} (\bibinfo{year}{1998}).
\newblock \textit{\bibinfo{title}{Statistical Learning Theory}}.
\newblock \bibinfo{publisher}{Wiley-Interscience}.
\bibitem[{Voigt and Bussche(2017)}]{Voigt2017}
\bibinfo{author}{Voigt, P}, \bibinfo{author}{Bussche, Avd}
  (\bibinfo{year}{2017}).
\newblock \textit{\bibinfo{title}{The EU General Data Protection Regulation
  (GDPR): A Practical Guide}}.
\newblock \bibinfo{edition}{1st} ed., \bibinfo{publisher}{Springer Publishing
  Company, Incorporated}.
\bibitem[{Vukovic et~al.(2012)Vukovic, Delibasic, Uzelac and
  Suknovic}]{VUKOVIC20128389}
\bibinfo{author}{Vukovic, S}, \bibinfo{author}{Delibasic, B},
  \bibinfo{author}{Uzelac, A}, \bibinfo{author}{Suknovic, M}
  (\bibinfo{year}{2012}).
\newblock \bibinfo{title}{A case-based reasoning model that uses preference
  theory functions for credit scoring}.
\newblock \textit{\bibinfo{journal}{Expert Systems with Applications}}
  \bibinfo{volume}{39}(\bibinfo{number}{9}): \bibinfo{pages}{8389--8395}.
\newblock \DOIprefix\doi{10.1016/j.eswa.2012.01.181}.
\bibitem[{Wihartiko et~al.(2018)Wihartiko, Wijayanti and
  Virgantari}]{Wihartiko2018}
\bibinfo{author}{Wihartiko, FD}, \bibinfo{author}{Wijayanti, H},
  \bibinfo{author}{Virgantari, F} (\bibinfo{year}{2018}).
\newblock \bibinfo{title}{Performance comparison of genetic algorithms and
  particle swarm optimization for model integer programming bus timetabling
  problem}.
\newblock \textit{\bibinfo{journal}{{IOP} Conference Series: Materials Science
  and Engineering}} \bibinfo{volume}{332}: \bibinfo{pages}{012020}.
\newblock \DOIprefix\doi{10.1088/1757-899x/332/1/012020}.
\bibitem[{Zheng et~al.(2019)Zheng, Li, Liu, Jia and Sheu}]{ZHENG2019227}
\bibinfo{author}{Zheng, XX}, \bibinfo{author}{Li, DF}, \bibinfo{author}{Liu,
  Z}, \bibinfo{author}{Jia, F}, \bibinfo{author}{Sheu, JB}
  (\bibinfo{year}{2019}).
\newblock \bibinfo{title}{Coordinating a closed-loop supply chain with fairness
  concerns through variable-weighted shapley values}.
\newblock \textit{\bibinfo{journal}{Transportation Research Part E: Logistics
  and Transportation Review}} \bibinfo{volume}{126}: \bibinfo{pages}{227--253}.
\newblock \DOIprefix\doi{https://doi.org/10.1016/j.tre.2019.04.006}.
\bibitem[{Zhuang et~al.(2009)Zhuang, Churilov, Burstein and
  Sikaris}]{Zhuang_EJOR}
\bibinfo{author}{Zhuang, ZY}, \bibinfo{author}{Churilov, L},
  \bibinfo{author}{Burstein, F}, \bibinfo{author}{Sikaris, K}
  (\bibinfo{year}{2009}).
\newblock \bibinfo{title}{Combining data mining and case-based reasoning for
  intelligent decision support for pathology ordering by general
  practitioners}.
\newblock \textit{\bibinfo{journal}{European Journal of Operational Research}}
  \bibinfo{volume}{195}(\bibinfo{number}{3}): \bibinfo{pages}{662--675}.
\newblock \DOIprefix\doi{https://doi.org/10.1016/j.ejor.2007.11.003}.
\bibitem[{Zien et~al.(2009)Zien, Kr{\"a}mer, Sonnenburg and
  R{\"a}tsch}]{Zien2009}
\bibinfo{author}{Zien, A}, \bibinfo{author}{Kr{\"a}mer, N},
  \bibinfo{author}{Sonnenburg, S}, \bibinfo{author}{R{\"a}tsch, G}
  (\bibinfo{year}{2009}).
\newblock \bibinfo{title}{The feature importance ranking measure}, in:
  \bibinfo{editor}{Buntine, W}, \bibinfo{editor}{Grobelnik, M},
  \bibinfo{editor}{Mladeni{\'{c}}, D}, \bibinfo{editor}{Shawe-Taylor, J}
  (Eds.), \bibinfo{booktitle}{Machine Learning and Knowledge Discovery in
  Databases}, \textit{\bibinfo{publisher}{Springer Berlin Heidelberg},
  \bibinfo{address}{Berlin, Heidelberg}}.  \bibinfo{pages}{694--709}.

\end{thebibliography}
\newpage
\section*{Appendix}
\appendix 
\renewcommand{\thesubsection}{\Alph{subsection}}
\renewcommand\thesection{\arabic{section}}
\renewcommand{\thetable}{\Alph{section}A.\arabic{table}}
\renewcommand{\thefigure}{\Alph{section}A.\arabic{figure}}
\renewcommand{\theequation}{\Alph{section}A.\arabic{equation}}
\subsection{Simple example of similarity calculation in CBR system}\label{Instance}
Assume there are a query company $Q$ and companies dataset $C$ which have twenty-eight ($L=28$) features, as shown in Table \ref{t:instance_1}. The bankruptcy of the query case is predicted based on the most similar companies from dataset $C$. First of all, we need to calculate all the similarities between $Q$ and each element $C_i$ of $C$. Each feature of companies has a weight (feature importance) and sum of all weights are equal to one ($\sum{\bm{w_{j}}} = 1$, $j=1,2, ..., 28$). For instance, the weights of Sales and Equity are 0.0494 ($\bm{w_{1}}$) and 0.0334 ($\bm{w_{2}}$), respectively. The parameters $\bm{a_{1}}$ and $\bm{b_{2}}$ for those two features in Equation (2) are 4.90 and 7.18, respectively. Then, the values of features in database are scaled to [0, 1] as shown in Table \ref{t:instance_2}. Thus, the difference between maximum and minimum value of each feature in database is equal to one ($D_j$ = 1). 
\begin{table}[ht!]
\caption{The features of query company $Q$ and database companies $C_i$}\label{t:instance_1}
 \begin{threeparttable}
 \begin{tabular*}{\textwidth}{p{3cm}  p{2.5cm} p{2.5cm} p{2cm}  p{2cm} P{2cm}} 
\toprule
Case &Sales&Equity&...&Net income&Bankrupt\\
\midrule
Company $Q$&4,233,270.93 &38,346.89&...&89,476.07&???\\
Company $C_1$&1,614,148.46 &40,903.35&...&14,316.17& Yes\\
Company $C_2$&1,801,792.58 &74,137.32&...&15,850.04& No\\
...&... &...&...&...&...\\

\bottomrule
\end{tabular*}
 \begin{tablenotes}
      \item Notes: based on most similar companies $C_i$ from database, the bankruptcy of company $Q$ is predicted.  
    \end{tablenotes}
  \end{threeparttable}
\end{table}

\begin{table}[ht!]
\caption{The scaled features of query company $Q$ and database companies $C_i$}\label{t:instance_2}
 \begin{threeparttable}
 \begin{tabular*}{\textwidth}{p{3cm}  p{2.5cm} p{2.5cm} p{2.cm}  p{2.cm} P{2cm}} 
\toprule
Case &Sales&Equity&...&Net income&Bankrupt\\
\midrule
Company $Q$&0.0175 ($q_1$) &0.0143 ($q_2$)&...&0.0235 ($q_{28}$)&???\\
Company $C_1$&0.0125 ($c_{1,1}$)&0.0153 ($c_{1,2}$)&...&0.0163 ($c_{1,28}$)& Yes \\
Company $C_2$&0.0132 ($c_{2,1}$)&0.0172 ($c_{2,2}$)&...&0.0168 ($c_{2,28}$)& No \\
...&... &...&...&...&...\\

\bottomrule
\end{tabular*}
 \begin{tablenotes}
      \item  
    \end{tablenotes}
  \end{threeparttable}
\end{table}

\noindent The similarity calculation of $Q$ and $C_1$ based on Equations (1) and (2) is given: 
\begin{align*}
\text{sim}_1(q_1,c_{1,1})&= \left\{\frac{D_{1}-(q_1-c_{1,1})}{D_{1}}\right\}^{\bm{a_{1}}} = \left\{\frac{1-(0.0175-0.0125 )}{1}\right\}^{\bm{4.90}} = 0.9757\\
\text{sim}_2(q_2,c_{1,2})&= \left\{\frac{D_{2}-(c_{1,2}-q_2)}{D_{2}}\right\}^{\bm{b_{2}}} = \left\{\frac{1-(0.0153-0.0143)}{1}\right\}^{\bm{7.18}} = 0.9928\\
\text{sim}_j(q_j,c_{1,j})&=...\ ... \\
\text{Sim}(Q,C_1) &= \sqrt{\sum_{j=1}^{28} \bm{w_{j}}\times\{\text{sim}_j(q_j,c_{1,j})\}^2} \\
&= \sqrt{\bm{0.0494}\times(0.9757)^2 + \bm{0.0334}\times(0.9928)^2 + ...} \\
&= 3.5224
\end{align*}
\newpage
\subsection{Feature relevance scoring methods}\label{score_methods}

A brief description of the six scoring methods are as follows:
\begin{enumerate}[label=(\arabic*)]
\item Gini \citep{Ceriani2012}: a typical metric used in decision trees to determine which characteristic should be used to split the current node for efficient decision tree creation. It is a measure of statistical dispersion and can be regarded as a measure of impurity for a characteristic or inequality among the values in a frequency distribution.

\item Information entropy \citep{Kullback59}: this metric measures the increase in information entropy using a characteristic of the class.

\item Mutual information \citep{Kraskov2004}: this method calculates the relationship between two variables and assesses the reduction in uncertainty for one variable when the value of the other variable is known.

\item Chi2 \citep{Cost1993}: this method examines each feature separately by calculating the chi-squared statistic for each class.

\item ANOVA \citep{LIN201164}: ANOVA stands for Analysis of variance. It is a statistical method used to check the means of two or more groups that are significantly different from each other. Here it ranks the features on the basis of their variance.

\item ReliefF \citep{Kononenko1997}: this method utilizes an attribute's ability to differentiate across classes on comparable data instances.
\end{enumerate}

\subsection{Parallel computing}\label{Parallel_computing}
Parallel computing is a type of computation where large calculations can be divided into smaller ones, and their computing processes are carried out simultaneously. The potential speedup of an algorithm on a parallel computing framework is given by Amdahl's law, which can be expressed mathematically as follows:
\begin{equation}
    \text{{Speedup}}= \frac{1}{(1-p)+\frac{p}{s}}
\end{equation}
\noindent where $\text{{Speedup}}$ is the theoretical maximum speedup of the execution of the whole task, and $p$ is the proportion of a system or program that can be made parallel and $s$ stands for the number of processors. 

One successful application of GPU-based parallel computing is deep learning, which is a typical intensive computing and training task that can be split. For the CBR querying process, it also can be paralleled. In particular, the similarity calculation between a query and each case can be processed simultaneously. The algorithm for predicting N queries with L features (query matrix) based on M cases with L features (reference matrix) is shown as follows:  \\ 

\begin{algorithm}[H]
\DontPrintSemicolon
  
\KwInput{N$\times$L query matrix, M$\times$L reference matrix, $\bm{a_j}$, $\bm{b_j}$, $\bm{w_j}$, $D_j$ and $\bm{k}$}
\KwOutput{N prediction vector $Prediction_{n}$}
\tcp{Each thread simultaneously calculates each similarity $\text{sim}_{n,j}$ between $q_{n,j}$ and $c_{m,j}$, where $n$ = 1, ..., N and $m$ = 1, ..., M.}
\While{calculate the similarity between $q_{n,j}$ and $c_{m,j}$}
  {
  \For{$j:= 1$ \KwTo L}
  {\eIf{$q_{n,j} \leq c_{m,j}$}
    {
        $\text{sim}_{n,j} = \left\{\frac{D_{j}-(c_{m,j} - q_{n,j} )}{D_{j}} \right\}^{\bm{a_j}}$     
    }
    {
    	$\text{sim}_{n,j} = \left\{\frac{D_{j}-( q_{n,j}-c_{m,j})}{D_{j}} \right\}^{\bm{b_j}}$ 
    }
    }
    Synthread() 
    \tcp{Wait for the computing completion for all the similarities $\text{sim}_{n,*}$.}
    ${\text{Sim}_{n}} = \sqrt{\sum_{j=1}^L w_j \times \text{sim}_{n,j}^2}$
    }
   \tcp{Wait for  the computing completion for all the similarities ${\text{Sim}_{n}}$ for $n_{th}$ query $q_{n,*}$}
   synchronized for query $q_{n,*}$.\;
   Sort and select the $\bm{k}$ most similar cases with $q_{n,*}$ from ${\text{Sim}_n},\ for\  n = 1, ..., N.$\;
   Voting $\bm{k}$ most similar cases to obtain the prediction for $q_{n,*}$: $Prediction_{n}$.\;
    
\caption{Similarity calculation pseudo code}
\end{algorithm}
\subsection{PSO}\label{PSO}
PSO is a computational technique for optimizing a problem by incrementally improving a solution as determined by a particular metric. The fundamental notion is that a population of particles traverses the search space. Each particle has knowledge about its current velocity, its own past best configuration ($\overrightarrow{p}(t)$), and the current global best solution ($\overrightarrow{g}(t)$). Based on this knowledge, each particle's velocity is adjusted to get it closer to both the current best solution and its previous best solution. The following equation guides how the velocity update is carried out:

\begin{equation}
\begin{split}
\overrightarrow{v}(t+1) & =\omega \overrightarrow{v}(t) + c_z{1}r_1(\overrightarrow{p}(t)-\overrightarrow{x}(t))+  c_{2}r_2(\overrightarrow{g}(t)-\overrightarrow{x}(t)) 
\end{split}
\end{equation}
\noindent where $c_1$ and  $c_2$ are constants defined beforehand, that determine the significance of $\overrightarrow{p}(t)$ and $\overrightarrow{g}(t)$. $\overrightarrow{v}(t)$ is the velocity of the particle, $\overrightarrow{x}(t)$ is the current particle position, $r_1$ and $r_2$ are random numbers from the interval [0,1], and $\omega$ is a constant ($0 \leq \omega \le 1$).
The new position is calculated by summing the previous position and the new velocity as follows:
\begin{equation}
\overrightarrow{x}(t+1) = \overrightarrow{x}(t) +\overrightarrow{v}(t+1)
\end{equation}

If the best individual solution in each iteration outperforms the best global solution, the best individual solution will be updated. Until a stopping requirement is met, this iterative process is repeated. PSO is employed in the proposed CBR system to look for the best settings for each feature similarity function.
\subsection{Measure metrics}\label{Measure_metrics}
TP (true positive) is the number of correctly classified positive instances. TN (true negative) is the number of correctly classified negative instances. FP (false positive) is the number of positive instances misclassified. FN (false negative) is the number of negative instances misclassified. 
\begin{enumerate}[label=(\arabic*)]

\item The equation of Accuracy can be described as follows:
\begin{equation}
    \text{Accuracy} = \frac{\text{TP} + \text{TN}}{\text{TP}+\text{FN}+\text{FP}+\text{TN}}
\end{equation}
\item The type I error is referred to as the FPR, which can be calculated as
\begin{equation}
    \text{FPR} = \frac{\text{FP}}{\text{TN}+\text{FP}}
\end{equation}
\item The type II error is referred to as the FNR, which can be calculated as
\begin{equation}
    \text{FNR} = \frac{\text{FN}}{\text{FN}+\text{TP}}
\end{equation}

\item Sensitivity or recall is referred to as the TPR, which can be calculated as
\begin{equation}
    \text{TPR} = \frac{\text{TP}}{\text{TP}+\text{FN}}
\end{equation}

\item Specificity or selectivity is referred to as TNR, which can be calculated as
\begin{equation}
    \text{TNR} = \frac{\text{TN}}{\text{FP}+\text{TN}}
\end{equation}
\item F-measure is the harmonic mean of precision and recall. In our study, the weighted average F-measure is used, which can obtained as 
\begin{equation}
    \text{F\mbox{-}measure} =  \frac{(1+\beta)^2\text{TP}}{(1+\beta)^2\text{TP}+\beta^2\text{FN}+\text{FP}}
\end{equation}
\item G-mean is the geometric mean of recall and precision, which can be calculated as
\begin{equation}
    \text{G\mbox{-}mean} = \sqrt{\text{TNR} \times \text{TPR}}
\end{equation}

\item AUC can be calculated as
\begin{equation}
    \text{AUC} = \frac{1+\text{TPR}-\text{FPR}}{2}=\frac{\text{TPR}+\text{TNR}}{2}
\end{equation}
\item MCC can be calculated as
\begin{equation}
    \text{MCC} = \frac{\text{TP}\times\text{TN}-\text{FP}\times\text{FN}}{\sqrt{(\text{TP}+\text{FP})(\text{TP}+\text{FN})(\text{TN}+\text{FP})(\text{TN}+\text{FN})}}
\end{equation}

\end{enumerate}
\subsection{Benchmark models}\label{Benchmark_models}
\noindent The benchmark models are briefly introduced as follows:\\

\noindent\underline{Logistic regression:}\\
Logistic regression is a mathematical modeling approach that can be used to describe the relationship of several variables to a dichotomous dependent variable \citep{Kleinbaum1994}. It is an efficient and powerful way to analyze the effect of a group of independent variables on a binary outcome \citep{Stoltzfus20111099}. In logistic regression, regularization is used to reduce generalization error and prevent the algorithm from overfitting in the feature-rich dataset. The Ridge and Lasso methods are most commonly used. Consequently, the inverse of regularization strength is also needed to determine. The smaller values specify stronger regularization. The best model can be found by cross-validation grid search. \\

\noindent\underline{$k$-nearest neighbor:}\\
$k$-nearest neighbors algorithm is a non-parametric classification method, which means it does not make any assumption on underlying data. It only considers the $k$ nearest neighbors to classify the query point \citep{Mitchell1997}. The hyperparameter required to decide is the $k$. The best model is achieved through a cross-validation procedure by using a grid search for the k.\\

\noindent\underline{Support vector machine:}\\
Support vector machine employs a linear model to implement nonlinear class boundaries by mapping input vectors nonlinearly into a high-dimensional feature space \citep{Vapnik1998}. The transformation is based on a kernel function. Thus, the choice of kernel function is important. The other two hyperparameters needed to determine are the regularisation and gamma. The former is the penalty parameter which tells SVM optimization how much error is bearable, and the latter defines how far influences the calculation of plausible line of separation. The hyperparameters of the best model are searched via cross-validation. \\

\noindent\underline{Decision tree:}\\
A decision tree is a map of the possible outcomes of a series of related choices, where each internal choice (node) denotes a test on an attribute, each branch represents an outcome of the test, and each leaf node holds a class label \citep{Song2015}. There are several hyperparameters required to tune. Impurity is used to determine how decision tree nodes are split. Information gain and Gini Impurity are commonly used. The maximum depth of the tree and the minimum number of samples required to be at a leaf node are also important to tune. Cross-validation grid search is applied to find the optimal model.\\
 
\noindent\underline{Gaussian Naive Bayes:}\\
Naive Bayes Classifiers are based on the Bayesian rule and probability theorems and have a strong assumption that predictors should be independent of each other \citep{Mitchell1997}. Gaussian naive Bayes classification is an extension of the naive Bayes method with an assumption that the continuous values associated with each class are distributed according to a Gaussian distribution. No hyperparameter tuning is required in Gaussian Naive Bayes. \\

\noindent\underline{Multi-layer perceptron:}\\
A Multi-layer preceptron (MLP) is a class of feedforward artificial neural network (ANN), which consists of at least three layers of nodes: an input layer, a hidden layer and an output layer \citep{GARDNER19982627}. It is a supervised non-linear learning algorithm for either classification or regression. MLP requires tuning a number of hyperparameters such as the number of hidden neurons, layers, and iterations. Hyperband algorithm is used for hyperparameters optimization \citep{li20181}.\\

\noindent\underline{Random Forest:}\\
Random Forest is an ensemble ML predictor that aggregates a set of weak classification and regression trees (CARTs) into a strong predictive model \citep{Breiman2001}. The training algorithm for random forests applies the general technique of bootstrap aggregating. The bootstrapping procedure means training each decision by randomly sampling a training data subset with replacement, which leads to a better model performance by decreasing the variance of the model. The predictions of a single tree are highly sensitive to noise in its training set, but the average of multiple de-correlated trees is not suffering from noise.\\

\noindent\underline{XGBoost:}\\
XGBoost is an implementation of gradient boosted decision trees with some improvements that help increase the performance and operate in parallel. Similar to RF, XGBoost is typically a tree ensemble model that combines weak-based learning tree models but achieves stronger learning ability, which is a set of CARTs. The main principle of the boosting model is to iteratively fit CARTs to training data, and all the CARTs are weighted according to their performance. The final result is the performance combination of the tree models. One of the most important differences between XGBoost and RF is that the XGBoost always gives more importance to functional space when reducing the cost of a model, while Random Forest tries to give more preferences to hyperparameters to optimise the model \citep{Bentejac2021}.

\subsection{Parameters}\label{para}
The parameters $a$ and $b$ of all the variables are shown in Table \ref{t:a_and_b}.
\begin{table}[ht!]
\caption{Parameters $a$ and $b$ of variables}\label{t:a_and_b}
 \begin{threeparttable}
 \begin{tabular*}{\textwidth}{p{1cm}p{4.75cm}P{0.4cm}P{0.4cm} p{1cm} p{4.75cm}P{0.4cm}P{0.4cm}  } 
\toprule
\toprule
No.&Feature&$a$&$b$&No.& Feature&$a$&$b$\\
\midrule
VAR1&Cash& 5.86 & 1.17&VAR15&Accounts payable (A.P.)&1.14 & 2.33 \\
VAR2&Inventories& 1.23 & 0.83& VAR16&Sales&2.12 & 5.53\\
VAR3&Current assets& 1.95 & 1.01&VAR17&Administrative expenses&1.18 & 6.53\\
VAR4&Tangible assets& 2.16 & 1.38&VAR18 &Amortization depreciation&0.71 & 1.04 \\
VAR5&Intangible assets& 7.08 & 2.70& VAR19&Interest expenses&1.53 & 1.45\\
VAR6&Total assets&4.25 & 2.41&VAR20&EBIT&4.43 & 1.71 \\
VAR7&Accounts receivable (A.R.)&3.57 & 6.47&VAR21&Operating income&0.65 & 1.69 \\
VAR8&Lands and buildings&2.68 & 0.69&VAR22&Net income&2.57 & 0.61 \\
VAR9&Equity& 1.70 & 1.24&VAR23&Increase inventories&1.21 & 0.34\\
VAR10&Shareholder loan&2.08 & 1.22&VAR24&Increase liabilities&1.13 & 1.10\\
VAR11&Accrual for pension liabilities&1.64 & 5.37&VAR25&Increase cash&4.88 & 5.00 \\
VAR12&Total current liabilities&1.08 & 2.00&VAR26&A.R. against affiliated companies& 5.11 & 4.55\\
VAR13& Total long-term liabilities &0.79 & 1.32&VAR27&A.P. against affiliated companies&0.78 & 1.68\\
VAR14&Bank debt&1.63 & 0.84&VAR28& Number of employees&2.32 & 0.75\\
\bottomrule
\bottomrule
\end{tabular*}
 \begin{tablenotes}
    \end{tablenotes}
  \end{threeparttable}
\end{table}
\newpage
\subsection{Financial variables}\label{t:fin_var}
Table A.4 shows that the financial variables of Gerry Webber in 2017, 2018 and 2019 were used for the insolvency prediction for 2018, 2019 and 2020, respectively.
\begin{center}
\begin{table}[htbp!]
\centering
\small
\caption{\centering Financial variables of Gerry Weber in 2017, 2018 and 2019.}\label{t:pred_results}
 \begin{threeparttable}
 \begin{tabular*}{0.65\textwidth}{ L{1.5cm} R{2.5cm}R{2.5cm}R{2.5cm} } 
\toprule
\toprule
    \multirow{2}{*}{Variable}&\multicolumn{3}{c}{Year}\\
    \cmidrule(lr){2-4} 
&2017 &2018 &2019 \\
\toprule
VAR1 &47,069.0 &-106,758.0 &166,092.0 \\
VAR2 &163,389.0 &87,978.0 &65,065.0 \\
VAR3 &270,765.0 &218,680.0 &236,471.0 \\
VAR4 &272,924.0 &90,152.0 &316,498.0 \\
VAR5 &229,890.0 &23,369.0 &20,136.0 \\
VAR6 &789,907.0 &374,890.0 &580,667.0 \\
VAR7 &70,798.0 &60,123.0 &44,477.0 \\
VAR8 &175,621.0 &73,631.0 &68,648.0 \\
VAR9 &412,749.0 &1,065.0 &121,442.0 \\
VAR10 &- &- &- \\
VAR11 &291,000.0 &152,000.0 &- \\
VAR12 &91,036.0 &310,875.0 &142,839.0 \\
VAR13 &286,123.0 &62,950.0 &316,386.0 \\
VAR14 &221,867.0 &268,523.0 &3,259.0 \\
VAR15 &51,858.0 &33,722.0 &14,090.0 \\
VAR16 &880,885.0 &330,512.0 &215,566.0 \\
VAR17 &- &- &- \\
VAR18 &62,961.0 &137,743.0 &46,770.0 \\
VAR19 &-7,404.0 &-1,641.0 &-8,462.0 \\
VAR20 &15,675.0 &-130,199.0 &130,172.0 \\
VAR21 &11,494.0 &-130,199.0 &130,172.0 \\
VAR22 &-782,000.0 &-244,501.0 &119,322.0 \\
VAR23 &-9,898.0 &-75,411.0 &-22,913.0 \\
VAR24 &-37,448.0 &219,839.0 &-168,036.0 \\
VAR25 &-16,404.0 &-153,827.0 &272,850.0 \\
VAR26 &- &- &- \\
VAR27 &- &- &- \\
VAR28 &6.88 &3.85 &3.33 \\
\bottomrule
\bottomrule
\end{tabular*}
 \begin{tablenotes}
      \small
      \item Notes: The unit is the thousand Euros or the thousand. The ``-'' stands for the value that is missing.
    \end{tablenotes}
  \end{threeparttable}
\end{table}
\end{center}

\end{document}